

\input phyzzx


\overfullrule=0pt

\def\M{{\cal M}}
\def\p{\partial}
\def\wt{\widetilde}
\def\L{{\cal L}}
\def\zero{{(0)}}
\def\wh{\widehat}
\def\al{{( a )}}

\def\bet{{( b )}}
\def\ten{{(10)}}

\def\R{{\cal R}}


\def\define#1#2\par{\def#1{\Ref#1{#2}\edef#1{\noexpand\refmark{#1}}}}
\def\con#1#2\noc{\let\?=\Ref\let\<=\refmark\let\Ref=\REFS
         \let\refmark=\undefined#1\let\Ref=\REFSCON#2
         \let\Ref=\?\let\refmark=\<\refsend}

\let\refmark=\NPrefmark

\define\BLUM
J. Blum, preprint EFI-94-04 (hep-th/9401133).

\define\WITTENSUSY
E. Witten, Nucl. Phys. {\bf B202} (1982) 253.

\define\MULLER
K. Muller, Phys. Lett. {\bf B177} (1986) 389.

\define\DISTLER
J. Distler and S. Kachru, Nucl. Phys. {\bf B413} (1994) 213 (hep-th/9309110);
preprints PUPT-1465 (hep-th/9406090), PUPT-1464 (hep-th/9406091),
and references therein.

\define\GPY
D. Gross, R. Pisarski and L. Yaffe, Rev. Mod. Phys. {\bf 53} (1981) 43.

\define\ATHIT
M. Atiyah and N. Hitchin, Phys. Lett. {\bf 107A} (1985) 21; Phil. Trans.
R. Soc. Lond. {\bf A315} (1985) 459; The Geometry and Dynamics of Magnetic
Monopoles, Princeton University Press (1988), and references therein.

\define\MACST
A. Dabholkar, G. Gibbons, J. Harvey and F.R. Ruiz, Nucl. Phys. {\bf B340}
(1990) 33; A. Dabholkar and J. Harvey, Phys. Rev. Lett. {\bf 63} (1989) 719.

\define\VAFAWITTEN
C. Vafa and E. Witten, preprint HUTP-94-A017 (hep-th/9408074).

\define\SEIWIT
N. Seiberg and E. Witten, preprints RU-94-52 (hep-th/9407087);
RU-94-60 (to appear).

\define\DUFFLU
M. Duff and J. Lu, Nucl. Phys. {\bf B347} (1990) 394.

\define\SHAT
S. Shatashvili and C. Vafa, preprint HUTP-94/A016 (hep-th 9407025).

\define\CAND
P. Candelas, G. Horowitz, A. Strominger and E. Witten, Nucl. Phys. {\bf
B258} (1985) 46.

\define\ASPIN
P. Aspinwall and D. Morrison, preprint DUK-TH-94-68 (hep-th 9404151)
and references therein.

\define\CERES
A. Ceresole, R. D'Auria and S. Ferrara, preprint CERN-TH.7384/94
(hep-th/9408036).

\define\SEGAL
G. Segal, to appear.

\define\WARD
R. Ward, Phys. Lett. {\bf 158B} (1985) 424.

\define\GIVEON
L. Girardello, A. Giveon, M. Porrati and A. Zaffaroni, preprint
NYU-TH-94/06/02 (hep-th/9406128).

\define\SHVAFA
B. Greene, A. Shapere, C. Vafa and S. Yau, Nucl. Phys. {\bf B337} (1990) 1.

\define\DUFFKH
M. Duff and R. Khuri, Nucl. Phys. {\bf B411} (1994) 473 (hep-th/9305142),
Nucl. Phys.  {\bf B418} (1994) 195 (hep-th/9311120).

\define\NARAIN
K. Narain, Phys. Lett. {\bf B169} (1986) 41.

\define\NSW
K. Narain, H. Sarmadi and E. Witten, Nucl. Phys. {\bf B279} (1987) 369.

\define\BOUND
A. Sen, Phys. Lett. {\bf B329} (1994) 217 (hep-th/9402032).

\define\GAUNTHAR
J. Gauntlett and J. Harvey, preprint EFI-94-36 (hep-th/9407111).

\define\FERRARA
S. Ferrara, C. Kounnas and M. Porrati, Phys. Lett. {\bf B181} (1986)
263; M. Terentev, Sov. J. Nucl. Phys. {\bf 49} (1989) 713.

\define\DIMRED
J. Maharana and J. Schwarz, Nucl. Phys. {\bf B390} (1993) 3
(hep-th/9207016);
S. Hassan and A. Sen, Nucl. Phys. {\bf B375} (1992) 103 (hep-th/9109038).

\define\HORMOORE
J. Horne and G. Moore, preprint YCTP-P2-94 (hep-th/9403058).

\define\FONT
A. Font, L. Ibanez, D. Lust and F. Quevedo, Phys. Lett. {\bf B249}
(1990) 35; S.J. Rey, Phys. Rev. {\bf D43} (1991) 526.

\define\GAUNT
J. Gauntlett, Nucl. Phys. {\bf B400} (1993) 103 (hep-th/9205008);
Nucl. Phys. {\bf B411} (1994) 443
(hep-th/9305068).

\define\GHL
J. Gauntlett, J. Harvey and J. Liu, Nucl. Phys. {\bf B409} (1993)
363.

\define\GIBMAN
G. Gibbons and N. Manton, Nucl. Phys. {\bf B274} (1986) 183.

\define\HARSTR
J. Harvey and A. Strominger, Comm. Math. Phys. {\bf 151} (1993) 221.

\define\SREV
A. Sen, preprint TIFR-TH-94-03 (hep-th/9402002) (to appear in
Int. J. Mod. Phys. A), and references therein.

\define\SLTZ
A. Sen, Int. J. Mod. Phys. {\bf A8} (1993) 5079 (hep-th/9302038).

\define\MONTOLIVE
C. Montonen and D. Olive, Phys. Lett. {\bf B72} (1977) 117;
P. Goddard, J. Nyuts and D. Olive, Nucl. Phys. {\bf B125} (1977) 1.

\define\OSBORN
H. Osborn, Phys. Lett. {\bf B83} (1979) 321.

\define\SCOL
A. Sen, Nucl. Phys. {\bf B404} (1993) 109 (hep-th/9207053)
, Phys. Lett. {\bf B303} (1993) 22 (hep-th/9209016),
Mod. Phys. Lett. {\bf A8} (1993) 2023 (hep-th/9303057);
J.Schwarz and A. Sen,
Phys. Lett. {\bf B312} (1993) 105 (hep-th/9305185).

\define\SCHSEN
J. Schwarz and A. Sen,  Nucl. Phys. {\bf B411} (1994) 35 (hep-th/9304154).

\define\MARSCH
N. Marcus and J. Schwarz, Nucl. Phys. {\bf B228} (1983) 145.

\define\WITTENOLIVE
E. Witten and D. Olive, Phys. Lett. {\bf B78} (1978) 97.

\define\WITTENTHETA
E. Witten, Phys. Lett. {\bf B86} (1979) 283.

\define\ROCVER
A. Giveon, E. Rabinovici and G. Veneziano, Nucl. Phys. B322 (1989) 167;
A. Shapere and F. Wilzcek, Nucl. Phys. B322 (1989) 669;
M. Rocek and E. Verlinde, Nucl. Phys. {\bf B373} (1992) 630
(hep-th/9110053);
A. Giveon and M. Rocek, Nucl. Phys. B380 (1992) 128 (hep-th/9112070).

\define\MURMAN
N. Manton and M. Murray, preprint DAMTP 94-57 (hep-th/9407102).

\define\MANTONBOUND
N. Manton, Phys. Lett. {\bf B198} (1987) 226.

\define\MARTINEC
E. Martinec, Phys. Lett. {\bf B171} (1986) 189.

\define\OLD
G. Gibbons and P. Ruback, Comm. Math. Phys. {\bf 115} (1988) 267;
N. Manton and B. Schroers, Annals. of Phys. {\bf 225} (1993) 290.

{}~\hfill \vbox{ \hbox{hep-th/9408083}\hbox{TIFR-TH-94-19}
}

\title{STRONG-WEAK COUPLING DUALITY IN THREE DIMENSIONAL STRING THEORY}

\author{Ashoke Sen\foot{e-mail addresses: sen@theory.tifr.res.in,
sen@tifrvax.bitnet}}

\address{Tata Institute of Fundamental Research, Homi Bhabha Road, Bombay
400005, India}

\abstract

Following an old result of Marcus and Schwarz we argue that in the heterotic
string theory compactified on a seven dimensional torus, the target space
duality group O(7,23;Z) and the strong-weak coupling duality transformations
combine into the group O(8,24;Z). We discuss symmetry of the combined
spectrum of elementary particles and solitons in the theory, and also show
that the existence of this symmetry predicts the number of harmonic forms
on the moduli space of periodic arrays of BPS monopoles in (3+1) dimensions.
Finally, we show that the O(8,24;Z) transformations relate the soliton
solutions of Dabholkar et. al. representing the fundamental string to
the soliton solutions of Greene et. al. representing `stringy cosmic
strings'.

\vfill\eject

\chapter{Introduction and Summary}

There is mounting evidence\con\MONTOLIVE\OSBORN\FONT\SCOL\SCHSEN
\SREV\BOUND\GAUNTHAR\VAFAWITTEN\SEGAL\noc
that the heterotic string theory compactified
on a six dimensional torus is invariant under an SL(2,Z) group of
transformations that acts non-trivially on the coupling constant of the
theory. Some of the consequences of this symmetry has been analyzed
in refs.\HORMOORE\GIVEON.
It is natural to ask if this symmetry is present for more
realistic compactification of the theory, say for compactification on
a Calabi-Yau manifold. At present there does not seem to be a way to
answer this question, since testing this symmetry requires computing
some observables ({\it e.g.} mass spectrum, partition function etc.)
exactly, {\it i.e.} without using
perturbation theory. There are certain
non-renormalization theorems\con\WITTENOLIVE\MARTINEC\MACST\noc
in the toroidal compactification of the
heterotic string theory that allows us to calculate some quantities in
the theory exactly, and hence test
the symmetry. Also, in the field theory limit, the partition function of
a twisted version of the theory is exactly calculable\VAFAWITTEN.
No such powerful non-renormalization theorem exists in
the $N=1$ supersymmetric compactification of the heterotic string theory.
$N=2$ supersymmetric compactifications provide an intermediate case,
and the recent results of Seiberg and Witten\SEIWIT\ might be useful
in analyzing the structure of duality transformations in heterotic
string theory compactified on $K3\times S^1\times S^1$\ASPIN.
(For a recent analysis of this problem see ref.\CERES.)

In the absence of a mechanism to test this symmetry in a realistic
compactification scheme, we can at least try to ask if this symmetry is
present in other (non-realistic) compactification of the theory. This
might give a clue to how this symmetry might depend on
the compactification scheme. With this in mind, we shall study the
duality group of heterotic string theory compactified on a seven
dimensional torus.
In the low energy limit,
this will result in a three dimensional supergravity
theory with eight local supersymmetries.

Such theories were analyzed long ago by Marcus and Schwarz\MARSCH.
The only massless bosonic fields in this theory are the spin two
(non-propagating)
graviton, and a set of scalar fields; this is due to the fact
that in three dimensions vector fields are dual to the scalar fields
and hence all vector fields can be traded in for scalar fields. It
was shown in Ref.\MARSCH\ that the scalar fields parametrize the
coset O(8,24)/O(8)$\times$O(24), and that the effective action is
invariant under O(8,24) transformation (see also \DUFFLU).
The duality group of
the full string theory is expected to be a subgroup of this O(8,24)
group. Of this, a subgroup O(7,23;Z) is already known to be a symmetry
of the theory, $-$ this is the standard target space duality group for
the compactification of the heterotic string on a seven dimensional
torus\ROCVER.
In this paper we shall argue that the duality group in fact is
much larger, $-$ O(8,24;Z), $-$ with a generic element  of O(8,24;Z) acting
non-trivially on the coupling constant of the theory.\foot{This was
already hinted in ref.\SCHSEN.}

An intuitive understanding of this large group of symmetries can be given
as follows.\foot{This argument has been developed independently by
E. Witten (private communication).}
The three dimensional theory can be regarded as a four
dimensional theory compactified on a circle. Since the strong-weak
coupling duality group SL(2,Z) does not act on space-time, we would
expect that this duality symmetry will remain unbroken even when we
compactify one of the three spatial directions. There are, however, seven
independent ways of regarding the three dimensional theory as a four
dimensional theory compactified on a circle, since any of the seven
compact directions could be regarded as the non-compact direction in the
original four dimensional theory before compactification. Thus the
three dimensional theory, besides possessing the usual O(7,23;Z) target
space duality symmetry, is expected to be invariant under seven
different SL(2,Z) symmetry groups.
The transformations belonging to these different groups do not commute
with each other, and are thus expected to generate a big non-abelian
discrete group.
As we shall see, the O(8,24;Z) group contains
the O(7,23;Z) group, as well as each of the
SL(2,Z) groups, as its subgroup.

If there is a general lesson to be learned from the analysis of this
paper, it is probably that for a generic string compactification we
should not expect the duality group to be a product of the target
space duality group and the coupling constant duality group. In
particular, for string theory compactified on a Calabi-Yau manifold\CAND,
or for more general (0,2) string compactification\DISTLER, the
duality transformations may mix the axion-dilaton field with the other
moduli fields in a non-trivial manner. Other interesting compactifications
of the theory to three dimensions\SHAT\ may also provide a good
laboratory for testing this phenomenon.

The paper is organized as follows. In sect. 2 we carry out the dimensional
reduction of the ten dimensional theory to three dimensions, and show
that the resulting theory is manifestly O(8,24) invariant. We also identify
an O(8,24;Z) subgroup of this which we claim to be an exact symmetry of the
full string theory. In sect.3 we show that various classical solutions in
the three dimensional theory, representing periodic arrays of monopoles
in the four dimensional theory, are related to elementary string states
via specific O(8,24;Z) transformations. We discuss the quantization of
some of these classical solutions in sect.4 and show that the O(8,24;Z)
symmetry of the three dimensional theory makes definite prediction about the
existence of harmonic forms in the moduli space of these solutions. In
sect.5 we show that two of the hitherto unrelated, but similar solutions
in string theory, $-$ the `stringy cosmic string' solution of ref.\SHVAFA,
and the `fundamental string' solution of ref.\MACST, $-$ are in fact
related by a $Z_2$ subgroup of the O(8,24;Z) group. In this section we
also suggest a way by which O(8,24;Z) invariance of the theory might
resolve the usual infra-red problems associated with charged particles
in three dimensions. In the three appendices we discuss various
properties of periodic arrays of BPS monopole solutions in (3+1)
dimensions, since these solutions provide most of the testing ground
for the O(8,24;Z) symmetry of the three dimensional theory.

\chapter{O(8,24) Invariant Effective Action}

We shall begin by discussing how the dynamics of massless fields in the
heterotic string theory, compactified on a seven
dimensional torus, is described by an O(8,24) invariant effective action.
This will also establish the relationship between the fields that
transform naturally under O(8,24), and the fields appearing from the
dimensional reduction of the $N=1$ supergravity theory in ten dimensions,
and will be useful to us later for studying the O(8,24;Z) transformation
properties of elementary string excitations and solitons.

We start with the $N=1$
supergravity theory coupled to $N=1$ super Yang-Mills theory in ten
dimensions, and dimensionally reduce the theory from ten to three
dimensions.
Since at a generic point in the moduli space only the abelian
gauge fields give rise to massless fields in three dimensions, it
is enough to restrict to the U(1)$^{16}$ part of the ten
dimensional gauge group. The ten dimensional action is given by,
$$\eqalign{
C \int d^{10}z \sqrt{ - G^{(10)}}\,
e^{-\Phi^{(10)}}\Big(R^{(10)} +G^{\ten MN}\p_M\Phi^\ten \p_N\Phi^\ten
\cr
- {1\over 12} H^{(10)}_{MNP}
H^{(10)MNP} - {1\over 4} F^{(10)I}_{MN} F^{(10)IMN}\Big),
\cr }
\eqn\eone
$$
where $G^{(10)}_{MN}$, $B^{(10)}_{MN}$, $A^{(10)I}_M$, and
$\Phi^{(10)}$ are ten dimensional metric, anti-symmetric tensor
field, U(1) gauge fields and the scalar dilaton field
respectively ($0\le M, N \le 9$, $1\le I\le 16$), and,
$$\eqalign{
F^{(10)I}_{MN} &= \p_M A^{\ten I}_N - \p_N A^{\ten I}_M \cr
H^{(10)}_{MNP} &= (\p_M B^\ten_{NP} -{1\over 2} A_M^{\ten I}
F^{\ten I}_{NP}) + \hbox{cyclic permutations of $M$, $N$, $P$}.\cr}
\eqn\etwo
$$
We have ignored the fermion fields in writing down the action
\eone.
$C$ is an arbitrary constant that we shall choose appropriately for later
convenience. This constant can always be absorbed into a shift in the
dilaton field.

The dimensional reduction of this theory can be carried out following
the procedure of refs.\FERRARA\DIMRED. We shall follow the notation of
ref.\SREV\ and introduce the
`three dimensional fields'  $\wh G_{mn}$, $\wh B_{mn}$, $\wh
A^I_m$, $\Phi$, $A_\mu^\al$, $g_{\mu\nu}$ and $B_{\mu\nu}$ ($1\le
m\le 7$, $0\le \mu\le 2$, $1\le a \le 30$) through the
relations
$$\eqalign{
& \wh G_{mn}  = G^\ten_{m+2,n+2}, \quad  \wh B_{mn}  =
B^\ten_{m+2, n+2}, \quad  \wh A^I_m  = A^{\ten I}_{m+2},
\cr
& A^{(m)}_\mu  = {1\over 2}\wh G^{mn} G^\ten_{n+2,\mu}, \quad
A^{(I+14)}_\mu = -({1\over 2} A^{\ten I}_\mu - \wh A^I_n
A^{(n)}_\mu), \cr
&  A^{(m+7)}_\mu = {1\over 2}
B^\ten_{(m+2)\mu} - \wh B_{mn} A^{(n)}_\mu + {1\over 2}\wh A^I_m
A^{(I+14)}_\mu, \cr
& G_{\mu\nu} = G^\ten_{\mu\nu} - G^\ten_{(m+2)\mu} G^\ten_{(n+2)\nu} \wh
G^{mn}, \cr
& B_{\mu\nu} = B^\ten_{\mu\nu} - 4\wh B_{mn} A^{(m)}_\mu
A^{(n)}_\nu - 2 (A^{(m)}_\mu A^{(m+7)}_\nu - A^{(m)}_\nu A^{(m+7)}_\mu),
\cr
& \Phi = \Phi^\ten - {1\over 2} \ln\det \wh G, \quad g_{\mu\nu}= e^{-2
\Phi} G_{\mu\nu}, \cr
& \quad \quad
\quad 1\le m, n \le 7, \quad
0\le \mu, \nu \le 2, \quad 1\le I \le 16.
\cr}
\eqn\ethree
$$
Here $\wh G^{mn}$ denotes the inverse of the matrix $\wh G_{mn}$.
We now combine the scalar fields $\wh G_{mn}$, $\wh B_{mn}$, and
$\wh A_m^I$ into an  $O(7,23)$ matrix valued scalar field $M$.
For this we regard $\wh G_{mn}$, $\wh B_{mn}$ and $\wh A^I_m$ as
$7\times 7$, $7\times 7$, and $7\times 16$ matrices
respectively, and $\wh C_{mn} = {1\over 2} \wh A^I_m
\wh A^I_n$ as a $7\times 7$ matrix, and define $M$ to be the
$30\times 30$ dimensional matrix
$$
M = \pmatrix{\displaystyle \wh G^{-1} & \wh G^{-1} (\wh B + \wh
C) & \wh G^{-1}\wh A \cr (-\wh B + \wh C) \wh G^{-1} & (\wh G
- \wh B +
\wh C) \wh G^{-1} (\wh G + \wh B + \wh C) & (\wh G -\wh B +\wh
C)\wh G^{-1} \wh A \cr  \wh A^T \wh G^{-1} &  \wh A^T \wh G^{-1}
(\wh G + \wh B +\wh
C) & I_{16} + \wh A^T \wh G^{-1} \wh A \cr }.
\eqn\efour
$$
satisfying
$$
M L M^T = L, \quad \quad  M^T=M, \quad \quad L =\pmatrix{0 & I_7
& 0  \cr I_7 & 0 & 0 \cr 0 & 0 & -I_{16}},
\eqn\efive
$$
where  $I_n$ denotes the $n\times n$ identity matrix.

The effective action that governs the dynamics of the massless
fields in the three dimensional theory is obtained by
substituting in Eq.\eone\
the expressions for the ten dimensional fields in
terms of the three dimensional fields, and taking all
field configurations to be independent of the internal coordinates. The
result is
$$\eqalign{
S &= {1\over 4} \int d^3 x \sqrt{-  g} \, \big[ R_g -
g^{\mu\nu}
\p_\mu \Phi \p_\nu\Phi -{1\over 12}e^{-4\Phi} g^{\mu\mu'} g^{\nu\nu'}
g^{\rho\rho'} H_{\mu\nu\rho} H_{\mu'\nu'\rho'} \cr
&\quad\quad  - e^{-2\Phi}g^{\mu\mu'} g^{\nu\nu'} F^\al_{\mu\nu} (LML)_{ab}
F^\bet_{\mu'\nu'} + {1\over 8} g^{\mu\nu} Tr (\p_\mu M L \p_\nu
M L) \big]\cr}
\eqn\esix
$$
where
$$\eqalign{
F^\al_{\mu\nu} &= \p_\mu A^\al_\nu - \p_\nu A^\al_\mu \cr
H_{\mu\nu\rho} &= (\p_\mu B_{\nu\rho} + 2 A^\al_\mu
L_{ a  b } F^\bet_{\nu\rho}) + \hbox{cyclic permutations of
$\mu$, $\nu$, $\rho$},\cr}
\eqn\edefhf
$$
and $R_g$ is the scalar curvature associated with the three
dimensional metric $g_{\mu\nu}$. In deriving this result we have
chosen the normalization constant $C$ in Eq.\eone\ in such a way that
we get an overall normalization factor of 1/4 in front of Eq.\esix.

The action given in eq.\esix\ is invariant under the O(7,23) transformation
$$
M\to \wt \Omega M \wt \Omega^T, \quad
A_\mu^\al \to \wt\Omega_{ab} A^\bet_\mu,
\quad g_{\mu\nu}\to g_{\mu\nu}, \quad B_{\mu\nu}\to B_{\mu\nu}, \quad \Phi
\to \Phi,
\eqn\esixa
$$
where $\wt\Omega$ is an O(7,23) matrix, {\it i.e.} a $30\times 30$ matrix,
satisfying,
$$
\wt\Omega L \wt\Omega^T = L.
\eqn\esixb
$$
In three dimensions the field $B_{\mu\nu}$ has no physical degrees of
freedom. Using this fact, we shall restrict ourselves to backgrounds
characterized by
$$ H_{\mu\nu\rho}=0. \eqn\eseven $$
The equations of motion of the gauge fields $A_\mu^\al$ now take the
form:
$$
\p_\mu \Big( e^{-2\Phi} \sqrt{-g}\, g^{\mu\mu'} g^{\nu\nu'} (ML)_{ab}
F^\bet_{\mu'\nu'}\Big) =0.
\eqn\eeight
$$
This allows us to define a set of scalar fields $\psi^a$ through the
relations
$$ \sqrt{-g} e^{-2\Phi} g^{\mu\mu'} g^{\nu\nu'} (ML)_{ab} F^\bet_{\mu'
\nu'} = {1\over 2} \epsilon^{\mu\nu\rho} \p_\rho \psi^a.
\eqn\enine $$
The Bianchi identity for $F^\al_{\mu\nu}$, $\epsilon^{\mu\nu\rho}
\p_\mu F^\al_{\nu\rho}=0$, now takes the form:
$$
D^\mu \Big( e^{2\Phi} (ML)_{ab}\p_\mu \psi^b \Big) =0,
\eqn\eninea
$$
where $D_\mu$ denotes the covariant derivative which preserves the metric
$g_{\mu\nu}$.

Let us now regard $\psi$ as a 30 dimensional column vector, and define a
new 32$\times$32 matrix $\M$ as follows
$$
\M = \pmatrix{M+e^{2\Phi}\psi \psi^T & -e^{2\Phi}\psi & ML\psi +{1\over 2}
e^{2\Phi} \psi (\psi^T L \psi) \cr
- e^{2\Phi} \psi^T & e^{2\Phi} & -{1\over 2} e^{2\Phi} \psi^T L \psi \cr
\psi^T LM +{1\over 2} e^{2\Phi}\psi^T (\psi^TL\psi) & -{1\over 2}
e^{2\Phi} \psi^T L \psi & e^{-2\Phi} + \psi^T LML \psi +{1\over 4}
e^{2\Phi}(\psi^T L \psi)^2 \cr}.
\eqn\eten
$$
It can be verified by straightforward algebraic manipulations that
the matrix $\M$ satisfies the relations
$$
\M^T =\M, \quad \M^T \L \M = \L,
\eqn\eeleven
$$
where $\L$ is a 32$\times$32 matrix
$$ \L =\pmatrix{ L & 0 & 0\cr 0 & 0 & 1\cr 0 & 1 & 0\cr}.
\eqn\etena $$
For $H_{\mu\nu\rho}=0$, the equations of motion derived from the action
\esix, together with eq.\eninea, can be shown to be identical to the
equations of motion derived from the action
$$
S = {1\over 4} \int d^3 x \sqrt{-g} \Big [ R_g + {1\over 8} g^{\mu\nu}
Tr(\p_\mu \M \L \p_\nu \M \L) \Big ].
\eqn\etwelve
$$
This action is manifestly invariant under the O(8,24) transformation
$$ \M \to \Omega \M \Omega^T, \quad g_{\mu\nu} \to g_{\mu\nu},
\eqn\efourteen
$$
where $\Omega$ is a 32$\times$32 matrix satisfying,
$$
\Omega \L \Omega^T = \L.
\eqn\efifteen
$$
This establishes the O(8,24) invariance of the low energy
effective field theory
describing the three dimensional string theory.

We shall now show that this O(8,24) symmetry of the action may be understood
in terms of the O(7,23) symmetry \esixa, and the SL(2,R) symmetry of the
four dimensional effective action discussed in Ref.\SREV. We regard the
three dimensional effective theory as the result of dimensional reduction of
the four dimensional effective theory to three dimensions. Then the three
dimensional effective theory should possess all the
symmetries of the four dimensional effective
theory, as long as they are not broken by compactification. In particular,
the SL(2,R) symmetry of the four dimensional effective action (whose
discrete subgroup SL(2,Z) may be identified with the S-duality group of the
full four dimensional theory) should induce an SL(2,R) symmetry of the three
dimensional theory.
There are many ways of regarding a three dimensional theory as a
compactified four dimensional theory; let us, for
definiteness, take the four dimensional theory to be the one obtained by
compactifying the directions 4-9. Then the
three dimensional theory is obtained
by compactifying the third direction of the four dimensional theory on a
circle. The action of the SL(2,R) group on various `four dimensional fields'
has been given in Ref.\SREV. From this one can find the transformation
laws of various ten dimensional fields, and hence, by using eq.\ethree,
\enine\ and \eten, the transformation laws of various `three dimensional
fields'. It  turns out that the SL(2,R) transformation of the `four
dimensional fields' $-$ $\lambda\to (a\lambda+b)/(c\lambda +d)$ $-$
generated by the
matrix $\pmatrix{a & b\cr c & d}$ with $ad-bc=1$, corresponds to the
following transformation on the `three dimensional fields':
$$\M\to \Omega \M \Omega^T, \eqn\fourcomponenew $$
where,
$$
\Omega=\pmatrix{a & 0 & 0 & 0 & 0 & b & 0\cr
0 & I_6 & 0 & 0 & 0 & 0 & 0\cr
0 & 0 & d & 0 & 0 & 0 & -c \cr
0 & 0 & 0 & I_6 & 0 & 0 & 0 \cr
0 & 0 & 0 & 0 & I_{16} & 0 & 0 \cr
c & 0 & 0 & 0 & 0 & d & 0\cr
0 & 0 & -b & 0 & 0 & 0 & a\cr}.
\eqn\fourcomptwonew
$$
It can be easily verified that $\Omega$ given above represents an
O(8,24) transformation.
Furthermore, one can also verify that the full O(8,24) group of transformations
may be generated as a combination of the O(7,23) transformation \esixa, and
the SL(2,R) transformation given above. Thus the O(8,24) symmetry of the
three dimensional effective action may be regarded as a consequence of the
O(7,23) symmetry, and the SL(2,R) symmetry of the four dimensional theory.

We shall argue in the next section
that an O(8,24;Z) subgroup of this is a symmetry
of the full string theory. In doing so, we must first identify an
appropriate O(8,24;Z) subgroup of the full O(8,24) group which we want to
show is the symmetry of the full theory.\foot{In the basis that we have
chosen, O(8,24;Z) matrices are {\it not} O(8,24) matrices with integer
entries.}
Let us first identify
the O(7,23;Z) subgroup of the O(8,24) that represents the usual target
space duality symmetry of the three dimensional string theory\ROCVER.
This subgroup
is generated by matrices of the form:
$$ \pmatrix{\wt\Omega & 0 & 0\cr 0 & 1 & 0 \cr 0 & 0 & 1\cr},
\eqn\esixteen
$$
where $\wt\Omega$ is an O(7,23) matrix satisfying the
condition that $L\wt\Omega L$ preserves the 30 dimensional Lorenzian,
even, self-dual lattice $\Lambda_{30}$, with metric $L$,
which represents the allowed values of the electric
charge vectors of elementary string states\NARAIN.
This specifies the O(7,23;Z) subgroup of O(8,24) completely.
For convenience, we shall choose the lattice $\Lambda_{30}$ as
the lattice containing the set of vectors
$$ \pmatrix{n_1 \cr \cdot \cr \cdot \cr n_{14} \cr \vec \xi},
\quad\quad n_i\in Z, \quad \vec\xi \in \Lambda_{E_8\times E_8},
\eqn\eseventeen
$$
where $\Lambda_{E_8\times E_8}$ is the root lattice of $E_8\times E_8$.
This means that in the absence of any background scalar field, the theory
represents the $E_8\times E_8$ heterotic string theory, compactified on
a seven torus, with each of the seven radii being equal to the self-dual
radius. In the normalization convention of ref.\SREV\ $\alpha'=16$,
and hence the self-dual
radius corresponds to $R=\sqrt{\alpha'}=4$. This also shows
that the normalization
factor of $1/4$ in eq.\etwelve\ is consistent with the normalization
factor of $1/32\pi$ that appears in front of the four dimensional
effective action of ref.\SREV, with the extra factor of $8\pi$ coming
from the length of the extra internal dimension. The compatibility of the
normalizations of the four and the three dimensional effective actions
will be
important for us in our later analysis, since we shall use the known
solitons in four dimensions to construct three dimensional solitons.
Note that by restricting the lattice $\Lambda_{30}$ in this way we do
not suffer from any loss of generality,
since any other choice of the lattice is equivalent to the lattice
$\Lambda_{30}$ described above, with an
appropriate choice of the background scalar fields\NARAIN\NSW.

We shall now specify the O(8,24;Z) subgroup of O(8,24) by prescribing a
32 dimensional lattice $\Lambda_{32}$ which is preserved by
the O(8,24;Z) transformations.
We take this lattice to be the collection of vectors
of the form:
$$ \pmatrix{ \vec\beta \cr m_1 \cr m_2 \cr}, \quad \vec\beta\in
\Lambda_{30}, \quad m_1, m_2 \in Z.
\eqn\eeighteen
$$
O(8,24;Z) subgroup of O(8,24) is defined to be the group of O(8,24) matrices
$\Omega$, satisfying the condition that $\L\Omega\L$ preserves the lattice
$\Lambda_{32}$. (Equivalently, $\Omega$ preserves the lattice
$\L\Lambda_{32}$.) In the next section
we shall discuss the possibility that this O(8,24;Z) subgroup is a
symmetry of the full string theory.

Using eqs.\eseventeen\ and \eeighteen\ we see that the lattice $\Lambda_{32}$
contains the set of vectors
$$ \pmatrix{n_1\cr \cdot \cr \cdot \cr n_{14}\cr \vec \xi\cr m_1\cr m_2},
\quad \quad n_i, m_i\in Z, \quad \vec\xi \in \Lambda_{E_8\times E_8}.
\eqn\edefltt
$$
In order to check that we are on the right track, we shall
now verify that the S-duality group SL(2,Z) of the four
dimensional theory is a subgroup of the O(8,24;Z) group that preserves
the above lattice.
To do this, we regard the 3-dimensional theory as a four dimensional theory
compactified on a circle as before,
taking the four dimensional theory to be the one obtained by
compactifying the directions 4-9.
{}From eqs.\fourcomponenew, \fourcomptwonew, we see that
the SL(2,Z) transformation generated by the
matrix $\pmatrix{p & q\cr r & s}$ with $ps-qr=1$,
$p, q, r, s \in Z$,  corresponds to an O(8,24)
transformation generated by:
$$
\Omega=\pmatrix{p & 0 & 0 & 0 & 0 & q & 0\cr
0 & I_6 & 0 & 0 & 0 & 0 & 0\cr
0 & 0 & s & 0 & 0 & 0 & -r \cr
0 & 0 & 0 & I_6 & 0 & 0 & 0 \cr
0 & 0 & 0 & 0 & I_{16} & 0 & 0 \cr
r & 0 & 0 & 0 & 0 & s & 0\cr
0 & 0 & -q & 0 & 0 & 0 & p\cr}.
\eqn\fourcomptwo
$$
It can be easily verified that $\Omega$ given above preserves the lattice
$\L\Lambda_{32}$ for $\Lambda_{32}$ defined by eq.\edefltt, and hence
represents an
O(8,24;Z) transformation. If, instead of regarding the 3 direction as one
of the four dimensions, we choose any of the other directions
4-9 as one of the four dimensions, the matrix $\Omega$ is modified
appropriately, but in each case represents an O(8,24;Z) transformation.

The three dimensional string theory is
manifestly O(7,23;Z) invariant. Thus establishing O(8,24;Z) invariance
of the theory basically amounts to showing that the original SL(2,Z)
symmetry of the four dimensional theory is not  destroyed when we
compactify one of the three space-like directions in this
theory.\foot{Recent results of ref.\VAFAWITTEN\ provide strong evidence
in favor of this.} In the next two sections
we shall try to identify some of the soliton states
in the string theory which are related to the elementary string excitations
via the above SL(2,Z) subgroup of O(8,24;Z) transformations.
The existence of these soliton states
is a necessary requirement for the O(8,24;Z) invariance of the theory.

\chapter{Elementary Particles and Solitons}

The charge quantum numbers of elementary string excitations are
characterized by a 30 dimensional vector $\vec\alpha\in\Lambda_{30}$.
The asymptotic value of the field strength $F^\al_{\mu\nu}$ associated
with such an elementary particle can be calculated using the procedure
outlined in ref.\SREV. The answer in polar coordinates is
$$ \sqrt{-g} F^{(a)0\rho} \simeq - {1\over 4\pi }
e^{2\Phi} M_{ab} \alpha^b , \eqn\enineteen $$
where $(\rho,\theta)$
are the polar coordinates of the two dimensional space, with origin at
the location of the elementary particle. Using eq.\enine, we get the
following asymptotic form of $\psi^a$:
$$
\psi^a \simeq - {\theta\over 2\pi} L_{ab} \alpha^b + \hbox{ constant}.
\eqn\etwenty
$$
Thus under $\theta\to \theta-2\pi$,
$\psi^a \to \psi^a + L_{ab} \alpha^b$.
This induces the following transformation of the matrix $\M$:
$$ \M \to \Omega_E (\vec\alpha) \M \Omega_E^T(\vec\alpha),
\eqn\etwentytwo $$
where,
$$ \Omega_E(\vec\alpha) = \pmatrix{I_{30} & - L \alpha & 0\cr
0 & 1 & 0\cr \alpha^T & -{1\over 2}\vec\alpha^2 & 1\cr}, \quad \quad
\vec\alpha^2 \equiv \alpha^T L \alpha.
\eqn\etwentythree
$$
One can easily verify that $\Omega_E(\vec\alpha)$ is an O(8,24) matrix.
Furthermore, since the lattice $\Lambda_{30}$ is even and self-dual with
respect to the metric $L$, we have $\vec\alpha^2\equiv
\alpha^T L \alpha$=even, and $\alpha^T
L \beta$=integer for $\vec\alpha, \vec\beta \in \Lambda_{30}$. Using
these relations one can verify that $\L\Omega_E(\vec\alpha)\L$ preserves
the lattice $\Lambda_{32}$, {\it i.e.} acting on a vector of the form
given in eq.\eeighteen, it produces another vector of the same form.
Thus $\Omega_E(\vec\alpha)\in O(8,24;Z)$. This shows that elementary
string excitations in three dimensions may be regarded as vortices, with
the matrix valued field $\M$ transforming by an O(8,24;Z) transformation
as we go around the vortex.

Of special interest are the elementary string excitations which are invariant
under half of the supersymmetry transformations, and hence saturate the
Bogomol'nyi bound\WITTENOLIVE.
These were analyzed in detail in refs.\SREV\GAUNTHAR.
Elementary string states with $\vec\alpha^2=-2$ and saturating the
Bogomol'nyi bound are 16 fold degenerate, whereas those with
$\vec\alpha^2=0$ and saturating the Bogomol'nyi
bound are 16$\times$24 fold degenerate.

At this stage we should point out that
due to infra-red logarithmic divergence in the electro-magnetic
contribution to the self energy, both the elementary string excitations and
the soliton masses are divergent in this theory. This does not contradict
the Bogomol'nyi formula, since
the asymptotic values of the scalar fields which appear in the expression
for the Bogomol'nyi bound, also diverge logarithmically.
Because of the divergent self-energies, it is more convenient to use the
criteria of unbroken supersymmetry rather than saturation of the Bogomol'nyi
bound to characterize these states, although we shall continue to refer to
them as states saturating the Bogomol'nyi bound. The presence of these
infra-red divergences puts the analysis in the three dimensional theory on
a less solid footing than the corresponding analysis in the four dimensional
theory. However, these infrared divergences affect the
elementary particles and solitons in an identical manner since they are
caused by massless scalars and abelian gauge fields, and, as we have seen
in sec.2, the dynamics of these fields is manifestly O(8,24;Z) symmetric.
Thus, whatever be the physical effect of these infra-red divergences,
we do not expect them to cause a violation of the O(8,24;Z) symmetry.
In view of this, we shall continue to analyze the spectrum of states in this
theory ignoring the infra-red divergences.
It will also be understood that when we relate the asymptotic behavior of
various fields with various charge quantum numbers, we shall be talking
of an asymptotic region where the logarithmic growth of various fields
have not reached a large enough value so as to invalidate the classical
analysis.

Let us now turn to the analysis of the non-singular soliton solutions in
the three dimensional string theory. A class of these solutions may be
constructed using an idea developed in ref.\GHL, namely by starting from
a periodic array of known monopole solutions in four dimensions along
one particular direction (say the 3 direction) with periodicity
$2\pi R_{(3)}$,
and reinterpreting it as a three dimensional solution, with the third
direction compactified on a circle of radius $R_{(3)}$. (As stated before,
here $R_{(3)}=4$.) There are two
classes of known non-singular
monopole solutions in four dimensions, the BPS monopole
and the $H$-monopole, and in both cases multiple monopole solutions are
known to exist. SL(2,Z) invariance of the four dimensional theory in
fact demands the existence of many other multi-monopole solutions, with each
monopole carrying  magnetic charge vector $\bar\beta$, for some
specific set of vectors $\bar\beta$ belonging to a 28-dimensional
self-dual Lorentzian lattice\SREV. In our analysis of this section we
shall assume the existence of solutions representing
periodic arrays of these monopoles; we
shall not, however, need to know the specific forms of these solutions.
\foot{In fact, construction of many of these solutions may require us
to take stringy effects into account.}
Quantization of these soliton solutions will be
discussed only for the periodic arrays of BPS monopole solutions, since
this is the case that is best understood. For this purpose, we have
discussed some of the properties of these solutions in appendices A-C.
(Quantization of even a single $H$-monopole solution is plagued by our
lack of understanding of the physics at large and small distance
scales\GAUNTHAR, although the classical multiple $H$-monopole solutions
can be constructed explicitly\GHL.)

Let us consider a magnetic monopole solution in four dimensions carrying
magnetic charge vector $\bar\beta$. ($\bar\beta^2=-2$ corresponds to BPS
monopoles, and $\bar\beta^2=0$ corresponds to $H$-monopole solutions.)
If $\bar A^{(\bar a)}_{\bar\mu}$ denotes the four dimensional vector fields
($1\le \bar a\le 28$, $0\le \bar \mu\le 3$), then the total magnetic
flux of the gauge field $\bar A^{(\bar a)}_{\bar\mu}$
associated with this monopole solution is given by $4\pi \bar L_{\bar a
\bar b}
\bar \beta^{\bar b}$\SREV, where,
$$ \bar L =\pmatrix{0 & I_6 & 0\cr I_6 & 0 & 0\cr 0 & 0 & -I_{16}\cr}.
\eqn\eextraone
$$
Let us now consider an array of these monopoles in the
3 direction separated by a distance $8\pi$ (which corresponds to a
radius of compactification 4, the self-dual radius). Then the total
flux per unit length coming out of a cylinder enclosing the 3 axis
will be ${1\over 2} \bar L_{\bar a \bar b}\bar \beta^{\bar b}$.
As has been argued in appendix A, asymptotically, the dependence of various
fields on the third direction can be ignored. Thus
if $(\rho, \theta)$ denote the
polar coordinates of the 2 dimensional space spanned by the 1,2 directions,
we get,
$$ \p_\theta \bar A^{(\bar a)}_3 \simeq - {1\over 4\pi}
\bar L_{\bar a \bar b}
\bar \beta^{\bar b}.
\eqn\eextratwo
$$
Thus $\bar A^{(\bar a)}_3$ increases
by ${1\over 2} \bar L_{\bar a \bar b}
\bar\beta^{\bar b}$ as $\theta\to \theta-2\pi$.

The relationship between four dimensional fields and the ten dimensional
fields was given in ref.\SREV. Hence from eq.\eextratwo\ we can
determine how the asymptotic values of different ten dimensional fields
transform under $\theta\to\theta-2\pi$. This,
in turn, can be used to calculate how the asymptotic values of
different three dimensional fields, defined through eq.\ethree, \efour,
\enine\ and \eten, transform under $\theta\to\theta-2\pi$.
We find that if we identify the direction $\bar \mu=3$
with the first compact direction of the 7 dimensional torus ($m=1$), and
if,
$$\bar \beta = -\pmatrix{\vec a_1 \cr \vec a_2 \cr \vec \xi\cr}, \quad
\quad \vec a_1, \vec a_2 \in Z^6, \quad \vec\xi\in \Lambda_{E_8\times E_8},
\eqn\eextrathree $$
then, as $\theta\to \theta-2\pi$,
$$ \M \to \Omega_S(\vec a_1, \vec a_2, \vec \xi) \M
\Omega_S^T(\vec a_1, \vec a_2, \vec\xi),
\eqn\etwentysixpr
$$
where,
$$
\Omega_S(\vec a_1, \vec a_2, \vec \xi)
=\pmatrix{1 & 0 & 0 & 0 & 0 & 0 & 0 \cr
a_2 & I_6 & 0 &  0 & 0 & 0 & 0 \cr
-{1\over 2}\bar\beta^2 & -a_1^T & 1 & -a_2^T & -\xi^T & 0 & 0\cr
a_1 & 0 & 0 & I_6 & 0 & 0 & 0 \cr
-\xi & 0 & 0 & 0 & I_{16} & 0 & 0\cr
0 & 0 & 0 & 0 & 0 & 1 & 0\cr
0 & 0 & 0 & 0 & 0 & 0 & 1\cr}.
\eqn\etwentysevenpr
$$
Here,
$$
\bar\beta^2\equiv \bar\beta^T\bar L \bar\beta = 2 \vec a_1\cdot \vec
a_2 - \vec \xi^2.
\eqn\etwentyeight
$$
It can be easily verified that $\Omega_S(\vec a_1, \vec a_2, \vec \xi)$
is an O(8,24;Z) matrix. Thus we see that these soliton solutions can
also be regarded as vortex solutions, with the matrix $\M$ transforming
by a non-trivial O(8,24;Z) transformation as we go around the vortex.
In fact, $\Omega_S$ represents an element of the target space duality
group O(7,23;Z). Although, unlike ordinary vortex solutions, these
solutions depend on the internal coordinate $x^3$, it has been argued
in appendix A that the $x^3$ dependence of various fields falls off
exponentially as we move away from the soliton core.

We shall shortly see that the elementary particles and the soliton
solutions are related by O(8,24;Z) transformations. However, we shall
first generalize the soliton solution by taking into account the electric
charge quantum numbers that the soliton is allowed to carry. First of
all, the four dimensional monopole solution can carry an electric charge
vector $-n\bar\beta$ for any integer $n$, this electric charge arises
from the quantization of an appropriate zero mode of the monopole
solution. This induces asymptotic values of $\p_\theta\psi^a$ as in
the case of electrically charged elementary string excitations. Furthermore,
the solution can also carry a momentum in the internal direction labelled
by $\bar \mu=3$. This momentum will be labelled by an integer $k$, and,
from the three dimensional viewpoint, will correspond to a new
charge quantum number associated with the three dimensional gauge fields.
Hence this will also induce an asymptotic value of $\p_\theta\psi^a$.
The net result is that for a three dimensional soliton labelled by
the quantum numbers $(\bar\beta, n, k)$,
$$ \M \to \Omega_S(\vec a_1, \vec a_2, \vec \xi; n, k) \M
\Omega_S^T(\vec a_1, \vec a_2, \vec\xi; n, k),
\quad \hbox{as} \,\, \theta\to \theta-2\pi,
\eqn\etwentysix
$$
where,
$$
\Omega_S(\vec a_1, \vec a_2, \vec \xi; n,k)
=\pmatrix{1 & 0 & 0 & 0 & 0 & 0 & 0 \cr
a_2 & I_6 & 0 &  0 & 0 & -na_2 & 0 \cr
-{1\over 2}\bar\beta^2 & -a_1^T & 1 & -a_2^T & -\xi^T & {1\over 2}
n\bar\beta^2 -k & 0\cr
a_1 & 0 & 0 & I_6 & 0 & -n a_1 & 0 \cr
-\xi & 0 & 0 & 0 & I_{16} & n\xi & 0\cr
0 & 0 & 0 & 0 & 0 & 1 & 0\cr
{n\over 2}\bar\beta^2 +k & na_1^T & 0 & na_2^T & n\xi^T & -{1\over 2}
n^2\bar\beta^2 & 1\cr}.
\eqn\etwentyseven
$$
It can be verified that $\Omega_S(\vec a_1, \vec a_2, \vec\xi;n,k)$
also represents an O(8,24;Z) matrix.

Thus we see that both, the elementary string excitations, and solitonic
states in the theory, saturating the Bogomol'nyi bound, are characterized
by O(8,24;Z) matrices $\Omega$, such that the matrix $\M$ transforms
to $\Omega \M \Omega^T$ as we go around the state once.
Let us now consider the effect of an O(8,24;Z) transformation $\omega$ on
such a state. Under such a transformation,
$$ \M \to \M'=\omega \M \omega^T. \eqn\ethirtyone $$
Thus, as $\theta\to \theta-2\pi$,
$$ \M' =\omega \M \omega^T \to \omega \Omega \M \Omega^T \omega^T
=\omega \Omega \omega^{-1} \M' (\omega^T)^{-1} \Omega^T \omega^T
=\Omega' \M' \Omega^{\prime T},
\eqn\ethirtytwo
$$
where,
$$
\Omega' = \omega \Omega \omega^{-1}.
\eqn\ethirtythree
$$
Thus under an O(8,24;Z) transformation generated by $\omega$, a state
characterized by the matrix $\Omega$ goes to a state characterized by
the matrix $\Omega'=\omega \Omega \omega^{-1}$.

We shall now show that the solitons characterized by the matrices
$\Omega_S(\vec a_1, \vec a_2, \vec \xi; n, k)$ given in eq.\etwentyseven\
are related to the
elementary string excitations characterized by the matrix
$\Omega_E(\vec\alpha)$ given in eq.\etwentythree\ through an O(8,24;Z)
transformation. This is easily verified by noting that
$$\Omega_S(\vec a_1, \vec a_2, \vec\xi; n, k) = \omega_n
\Omega_E(\vec\alpha_0) \omega_n^{-1},
\eqn\ethirtyfive
$$
where,
$$
\vec \alpha_0 = \pmatrix{k\cr \vec a_1 \cr 0 \cr \vec a_2\cr \vec\xi\cr},
\quad \quad \vec\alpha_0^2\equiv \alpha_0^T L \alpha_0 = 2 \vec a_1\cdot
\vec a_2 - \xi^2,
\eqn\ethirtyseven
$$
and,
$$
\omega_n= \pmatrix{n & 0 & 0 & 0 & 0 & -1 & 0\cr 0 & I_6 & 0 & 0 & 0
& 0 & 0\cr 0 & 0 & 0 & 0 & 0 & 0 & -1\cr 0 & 0 & 0 & I_6 & 0 & 0 & 0 \cr
0 & 0 & 0 & 0 & I_{16} & 0 & 0\cr 1 & 0 & 0 & 0 & 0 & 0 & 0\cr
0 & 0 & 1 & 0 & 0 & 0 & n\cr}.
\eqn\ethirtynine
$$
This shows that the soliton solutions are related to the elementary
excitations in string theory via an O(8,24;Z) transformation.\foot{This
result is not surprising, since in the limit of large radius of the
internal direction $x^3$, both the elementary particles and the solitons
can be identified to the corresponding states in four dimensions, which
are known to be related by SL(2,Z) transformations. But as we shall see
in the next section, O(8,24;Z) symmetry makes non-trivial predictions
about the moduli space of multi-soliton solutions.}
Note that
in constructing the soliton solutions we have treated the internal
direction $m=1$ as special, since that is the direction we
have identified as one of the directions in the four dimensional theory.
Other soliton
solutions in the theory, corresponding to any other internal direction
being identified as the fourth
direction may be constructed in a similar manner.
These solitons will also be related to the elementary string excitations
by appropriate O(8,24;Z) transformation.

So far we have only shown that the quantum numbers of the solitons
are related to those of elementary particles via an O(8,24;Z)
transformation. However, O(8,24;Z) invariance also requires
that the degeneracy of
states is identical for the elementary particles and solitons.
In order to investigate this question,
we need to quantize the collective coordinates of the soliton
solutions. As mentioned at the beginning of this discussion,
this will be done only for the periodic arrays of BPS monopole
solutions (characterized by
$\bar\beta^2=-2$). From the relationship between $\vec\alpha_0$ and
$\bar \beta$ given in eqs.\eextrathree, \ethirtyseven\ it follows that the
corresponding elementary string states have $(\vec\alpha_0)^2=-2$, and
hence are 16 fold degenerate\SREV. On the other hand, the moduli space of
periodic array of single BPS monopoles is 4 dimensional, and is given
by $R^2\times S^1 \times S^1$, with $R^2$ representing the location
of the array in the non-compact directions, the first $S^1$ denoting
the location of the array in the compact direction, and the second
$S^1$ representing the coordinate conjugate to the U(1) electric
charge.\foot{As is typical for solitons in three dimensions\WARD,
the zero modes associated with $R^2\times S^1\times S^1$ are
not strictly normalizable due to a logarithmic divergence from large
distances. This divergence is related to the fact that these
solitons have infinite
mass due to  logarithmic divergence in the integral of energy density from
large distance scale, and also affects the elementary string states.
A possible infra-red regularization that gives rise to normalizable zero
modes associated with the $R^2\times S^1\times S^1$ factor has been discussed
in appendix C.}
The quantum numbers $k$ and $n$ can be identified to momenta conjugate to
these two angular coordinates.
The structure of the fermionic zero modes is identical to that of a
single BPS monopole in 4 dimensions with moduli space $R^3\times S^1$,
since the supersymmetry algebra,
as well as the number of unbroken supersymmetry generators is the same
in the two cases. Standard analysis\OSBORN\GAUNT\ then shows that the
soliton state is 16-fold degenerate, in agreement with the requirement
of O(8,24;Z) symmetry.

In our discussion relating the solitons to elementary excitations,
we have considered only those elementary
string states which do not carry any winding number in the $x^3$ ($m=1$)
direction. This can be seen from eq.\ethirtyseven. Strings winding
along the $x^3$ direction will play a crucial role in the analysis
of section 5.

\chapter{Predictions of O(8,24;Z) Invariance}

As in the case of four dimensional theory, existence of O(8,24;Z)
symmetry predicts the existence of many new soliton states in the theory.
We shall consider only some specific examples here. Let us consider
the O(8,24;Z) matrix
$$
\omega=\pmatrix{p & 0 & 0 & 0 & 0 & q & 0\cr
0 & I_6 & 0 & 0 & 0 & 0 & 0\cr
0 & 0 & s & 0 & 0 & 0 & -r \cr
0 & 0 & 0 & I_6 & 0 & 0 & 0 \cr
0 & 0 & 0 & 0 & I_{16} & 0 & 0 \cr
r & 0 & 0 & 0 & 0 & s & 0\cr
0 & 0 & -q & 0 & 0 & 0 & p\cr}, \quad \quad ps -qr =1.
\eqn\eforty
$$
Applying this O(8,24;Z) transformation on an elementary string excitation
characterized by the matrix $\Omega_E(\vec\alpha_0)$, with $\vec\alpha_0$
given by eq.\ethirtyseven, we get the following O(8,24;Z) matrix:
$$
\omega \Omega_E(\vec\alpha_0) \omega^{-1}
= \pmatrix{1 & 0 & 0 & 0 & 0 & 0 & 0 \cr
r a_2 & I_6 & 0 & 0 & 0 & -p a_2 & 0 \cr
-{r^2\over 2}\vec\alpha_0^2 & -r a_1^T & 1 & -r a_2^T & - r\xi^T &
-k +{pr\over 2} \vec\alpha_0^2 & 0\cr
ra_1 & 0 & 0 & I_6 & 0 & -pa_1 & 0 \cr
-r\xi & 0 & 0 & 0 & I_{16} & p\xi & 0\cr
0 & 0 & 0 & 0 & 0 & 1 & 0 \cr
k + {pr\over 2}\vec\alpha_0^2 & p a_1^T & 0 & p a_2^T & p \xi^T &
-{p^2\over 2} \vec\alpha_0^2 & 1\cr}.
\eqn\efortyone
$$
Reversing the analysis that led to eq.\etwentyseven, we can easily verify
that a state characterized by the above O(8,24;Z) matrix corresponds to a
periodic array of $r$
monopoles in four dimensions, carrying electric and magnetic
charge quantum numbers $ -p \bar \beta $ and $r \bar \beta$ respectively,
and carrying $k$ units of momentum in the internal
direction $\bar \mu =3$. For $\bar\beta^2=-2$, {\it i.e.} for
BPS monopoles, O(8,24;Z) invariance again predicts a 16
fold degeneracy of these states.

In order to verify
this prediction, one may proceed as follows. First one assumes
that the scale of breaking of the SU(2) symmetry that leads to the construction
of the BPS monopoles, as well as the inverse radius of the
third direction, is small compared to the Planck mass. In this limit we can
ignore the gravitational effects, and regard the solitons as periodic array
of monopole solutions in $N=4$ supersymmetric SU(2) gauge theories in four
dimensions, with the SU(2) symmetry spontaneously broken by the vacuum
expectation value of the Higgs field. In particular, in this limit the
off-diagonal components $G^\ten_{3\mu}$ of the gravitational field, which
would behave like a gauge field in three dimensions and would couple
to the charge quantum number $k$, is ignored, since the corresponding gauge
coupling constant is proportional to the inverse radius of the third
direction measured in Planck units. The question of existence
of states characterized by eq.\efortyone\ now becomes
the question of whether the
SL(2,Z) symmetry of the N=4 supersymmetric Yang-Mills theory in
four dimensions survives under the
compactification of one of the space directions. Although the recent results
of Vafa and Witten\VAFAWITTEN\ show that the SL(2,Z) invariance
of the theory does not depend on the underlying four manifold, it will be
reassuring to directly verify the existence of these states. We shall now
discuss how this might be done.

As in the case of multi-monopole solutions,
the existence of
supersymmetric states representing multiple periodic arrays of monopoles,
and carrying multiple units of electric charge and momentum in the 3
direction,
can be studied by quantizing the collective coordinates
parametrizing the moduli space
of these solutions.
The moduli space of $r$ BPS monopoles
arranged in a periodic array has been discussed in some detail in appendix
B. As explained there, this moduli space is expected to have the structure
\foot{It has been shown in appendix C that even though the zero modes
associated with the coordinates on $R^2\times S^1\times S^1$ are not
normalizable, the zero modes associated with the coordinates on $\M_r^0$
are renormalizable.}
$$\M_r = R^2\times (S^1 \times S^1 \times \M^0_r/Z_r\times Z_r)
\eqn\structure $$
where
$\M^0_r$ is a $4(r-1)$ dimensional manifold.
The first $Z_r$ acts on the coordinate $X^3$ of the first $S^1$ as
$X^3\to X^3-(2\pi/r)$, and the second $Z_r$ acts on the coordinate
$\Theta$ of the second $S^1$ as $\Theta\to \Theta-(2\pi/r)$.
Both the $Z_r$ transformations also have non-trivial action on $\M^0_r$.
Physically,
for well separated arrays, the action of the first $Z_r$ on $\M_r$
corresponds to translating one of the periodic arrays by one
period, whereas the action of the second $Z_r$ on $\M_r$
corresponds to changing the U(1) phase of all the monopoles in one array
by $2\pi$.
Since $p$ and $k$ are the quantum numbers representing the total U(1)
charge, and the total momentum in the third direction respectively,
for a given value of $p$ and $k$ the $X^3$ and $\Theta$ dependent part
of the wave-function has the form $\exp(i k X^3 + i p \Theta)$.
Hence this part of the
wave-function picks up a factor of $\exp(-2\pi i k /r)$ under the
first $Z_r$ transformation, and a factor
of $\exp(-2 \pi i p/r)$ under the second $Z_r$ transformation.
Thus the part of the wave-function coming from
$\M^0_r$ must pick up a factor of $\exp(2\pi i k/r)$ under the action of
the first $Z_r$ and a factor of $\exp(2\pi i p/r)$ under the action of
the second $Z_r$.

As was shown in ref.\WITTENSUSY, and discussed in the specific context of
monopoles in refs.\GAUNT\BLUM\BOUND,
the supersymmetric states of the soliton,
saturating the Bogomol'nyi bound, are represented by harmonic forms
on the space $\M^0_r$. For each such harmonic form, we get a 16 fold
degenerate state from quantization of the supersymmetric partners of the
coordinates on $R^2\times S^1\times S^1$. From
the relation $ps -qr=1$ it follows that the numbers $p$ and $r$ are
relatively prime, but there is no restriction on $k$. Thus if we want
a 16 fold degenerate state for each of these values of $p$ and $k$, we
must demand that,

\noindent{\it For every integer $p$ that is relatively prime with
respect to $r$, and for every integer $k$, the space $\M^0_r$ has
a unique (and hence (anti-)self-dual) harmonic form, which picks up
a phase $\exp(2\pi i k/r)$ under the action of the first $Z_r$ group,
and a phase $\exp(2\pi i p/r)$ under the action of the second
$Z_r$ group.}

At present we do not have a direct proof of the above conjecture
for values of $r\ge 2$. We shall now give an indirect argument for
the existence of these harmonic forms starting from the corresponding
results on the $r$-monopole moduli space\OLD\BOUND\SEGAL.
Before we proceed we should, however, warn the reader that this
argument is based on various assumptions that look physically
plausible, but it does not constitute a rigorous proof.
For simplicity, we shall consider the case $r=2$, the extension to
the more general case is straightforward. Let us consider the case
when the radius of compactification $R_{(3)}$ of the third direction
is much larger than the scale of breaking of the SU(2) symmetry (which
we shall denote by $K$). Let $\R$ denote the region of $\M_2^0$ where
two monopoles are within a distance of order $K$ of each other, and
the resulting configuration is repeated along the third direction with
periodicity $2\pi R_{(3)}$. Let us, for definiteness, assume that
the 2-monopole clusters are located around the points $2\pi n R_{(3)}$
on the $3$-axis. In this region, the metric on $\M_2^0$ is close to
the metric on the two monopole relative moduli space $M_2^0$, since
for an infinitesimal motion in $\M_2^0$, the change in various fields
near the monopoles in one cluster comes mostly due to the monopoles
in that cluster. Contribution from the other clusters are suppressed
by inverse powers of $R_{(3)}$. Now, $M_2^0$ is known\OLD\BOUND\SEGAL\
to have a self-dual harmonic form $\Omega$ which picks up a phase of
$-1$ under the $Z_2$ action associated with the U(1) phase. $\Omega$
falls off exponentially as the two monopoles are pulled away from
each other to a distance much larger than $K$. This allows us to
construct an approximate harmonic 2-form $\wh\Omega$
on $\M_2^0$ which is equal to $\Omega$ in the region $\R$, and
falls off rapidly in the region of moduli space where we pull the
two monopoles in the cluster away from each other at a distance
$>>K$.

Let $h$ denote the action of the first $Z_2$ transformation on $\M_2^0$.
As explained in appendix B, this corresponds to moving one of the
monopoles in the cluster by a distance $\pi R_{(3)}$ along the positive
3-axis, and the other monopole by a distance $-\pi R_{(3)}$ along
the negative 3-axis. Let $h\circ \wh \Omega$ denote the action
of the diffeomorphism $h$ on the approximate harmonic form
$\wh \Omega$. Since the diffeomorphism generated by $h$ is a symmetry
of the metric, $h\circ \wh \Omega$ is also an approximate
harmonic form on $\M_2^0$. Furthermore, since $\wh \Omega$ has
support mostly in the region of $\M_2^0$ where the monopoles
cluster around the points $2\pi n R_{(3)}$, and (as follows from the
definition of $h$) $h\circ \wh \Omega$ has support mostly in
the region of $\M_2^0$ where the monopoles cluster around the
points $(2n +1)\pi R_{(3)}$, there is no significant overlap between
the forms $\wh \Omega$ and $h\circ \wh \Omega$. It then follows
that
$$ \omega^{(k)}\equiv \sum_{l=0}^1 e^{i \pi kl} h^l \circ \wh
\Omega \eqn\harmonic $$
are also approximate harmonic forms on $\M_2^0$ with the property that
$\omega^{(k)}$ picks up a phase of $\exp(i\pi k)$ under the action
of the first $Z_2$ group of transformations generated by $h$, and a
phase of $-1$ under the second $Z_2$ group of transformations
associated with the U(1) phase. The construction given in eq.\harmonic\
gives two independent $\omega^{(k)}$, which can be taken to be
$\omega^{(0)}$ and $\omega^{(1)}$.

Normally one would expect  that the above approximate harmonic forms
on $\M_2^0$ can be suitably modified so as to give exact eigenforms
$\bar\omega^{(k)}$ of the Laplacian with a small, but not
necessarily zero, eigenvalue $\lambda^{(k)}$.
Physically this effect may be understood by regarding the system as
a one dimensional periodic array of two monopole bound states (with
finite binding energy).
We expect the bound state energy to shift slightly due to the influence
of the other monopoles and the periodicity requirement, but do not
expect the state to disappear completely.
We shall now argue that in the present case
$\lambda^{(k)}$ vanishes exactly. This is proved by noting that if
$\lambda^{(k)}\ne 0$, then either $d$ or $\delta\equiv * d *$ (or both)
does not annihilate $\bar\omega^{(k)}$. Let us assume that $d
\bar\omega^{(k)}\ne 0$; the case where $\delta\bar\omega^{(k)}\ne 0$
can be dealt with in an identical manner.
We can then define a normalized differential form
$\wt\omega^{(k)} \equiv d \bar\omega^{(k)}/\sqrt{\lambda^{(k)}}$ which
is also an eigenform of the laplacian with eigenvalue $\lambda^{(k)}$.
Since for $R_{(3)}\to \infty$, $\lambda^{(k)}\to 0$, in this limit
$\wt\omega^{(k)}$ corresponds to a harmonic form on $\M_2^0$, and its
restriction to $\R$ would correspond to a new
harmonic form on the two monopole moduli space $M_2^0$.\foot{We are
implicitly assuming that $\wt\omega^{(k)}$ will also
have its support mostly inside $\R$ and $h(\R)$. This is physically
plausible, since bound states of monopoles with finite binding energy
are expected to have support only in regions where the monopoles are
close to each other.}  (Note that
in this limit $\wt\omega^{(k)}$ ceases to be an exact form, since
$d\bar\omega^{(k)}\to 0$.) But we
know from the analysis of the two monopole moduli space that such
harmonic forms do not exist\SEGAL.
This shows that our initial assumption
must be wrong, i.e. $\lambda^{(k)}$ must vanish.

This argument shows the existence of the relevant harmonic forms
$\bar\omega^{(k)}$ on
$\M_2^0$ for large $R_{(3)}$ and hence the existence of the relevant
supersymmetric state in the two soliton sector. It is probably clear
to the reader that the argument presented here is simply a variation
of the argument\WITTENOLIVE\ that once $\lambda^{(k)}$ is non-zero,
the number of states increase, and there is no convenient way of
getting rid of these extra states in the $R_{(3)}\to \infty$ limit.
We can now use the same continuity argument
as in ref.\WITTENOLIVE\ to argue that these states, and hence these
harmonic forms must exist also for finite values of $R_{(3)}$.
Alternatively, one could argue that the number of square integrable
harmonic forms on a manifold is a topological property of the
manifold, and is not expected to change under smooth deformations
of the manifold.

As in the case of four dimensional theories\SREV\BOUND, we can also give a
plausibility argument for the existence of these bound
states based on triangle
inequality.
Since the Bogomol'nyi formula gives infinite mass of all the states,
this is not a useful starting point. Instead, we shall base our calculation on
the direct analysis of the force between far away solitons. For this, let us
consider two arrays, one carrying $r_1$ units of magnetic charge and $p_1$
units of electric charge per unit period in the third
direction, and the other
carrying $r_2$ units of magnetic charge and $p_2$ units of electric charge
per unit period in the third direction. Using the results of ref.\MULLER\
giving the asymptotic Higgs and gauge fields around a dyon,
one can show that the  attractive force
(proportional to the inverse separation) due the exchange of massless Higgs
between the two arrays is
proportional to $\sqrt{(p_1)^2+(r_1)^2} \sqrt{(p_2)^2+(r_2)^2}$, whereas the
repulsive electric and magnetic force between the two arrays is proportional
to $(p_1p_2+r_1r_2)$.\foot{For simplicity, we have assumed that the coupling
constant is unity,
but this is not a necessary assumption.}
These two forces balance
each other only when the two vectors $\pmatrix{p_1\cr r_1}$ and
$\pmatrix{p_2\cr r_2}$ are proportional to each other. Otherwise the
attractive force due to the higgs exchange is always larger then the repulsive
electric and magnetic force.

Let us now consider a state carrying $p$ units of magnetic charge and $r$ units
of electric charge per unit period in the three direction. If $p$ and $r$ are
relatively prime, it is clear that the vector $\pmatrix{p\cr r}$ cannot be
written as a sum of two other vectors with integer entries which are
proportional to each other. In other words, if we try to regard the array
carrying the above magnetic and electric charges as a combination of two
different arrays, then for large separation, these two arrays experience a net
attractive force. It is then plausible that there are bound states of these
two arrays, and that there exists specific bound states which attain the
lowest possible energy, {\it i.e.} saturate the Bogomol'nyi bound. These
are precisely the states that are needed for establishing duality invariance
of the spectrum.

\chapter{Fundamental Strings and Stringy Cosmic Strings}

There are two kinds of different string like solutions that have been
constructed in four dimensional string theory.
The first kind, known as fundamental string,
describes the classical field configuration around a fundamental string and
was constructed in ref.\MACST. The second kind, which was called `stringy
cosmic string', and was constructed in ref.\SHVAFA\DUFFKH,
represents a solution
where one of the 6 compact directions decompactify at the core of the
string. Each of these two solutions may be regarded as a solution in the
three dimensional string theory by taking the direction, along which the
string extends, to be compact. In this section we shall show that these
two different classes of three dimensional solutions are, in fact, related
by an O(8,24;Z) transformation. Unlike in the previous section, our
discussion in this section will be based on the study of the classical
solutions themselves, but we expect that this can be elevated to the
quantum level once we properly quantize the collective coordinate excitations
of the corresponding solutions.

We begin with a discussion of the fundamental string solution.
In the notation of this paper, the field configuration
associated with a fundamental string solution, winding
once in the three direction and without carrying any other charges,
is given by,
$$\M = \pmatrix{{|\lambda|^2
\over \lambda_2} & 0 & 0 & 0 & 0 & {\lambda_1\over \lambda_2}
& 0 \cr
0 & I_6 & 0 & 0 & 0 & 0 & 0 \cr 0 & 0 & {1\over \lambda_2} & 0 & 0
& 0 & -{\lambda_1\over \lambda_2} \cr 0 & 0 & 0 & I_6 & 0 & 0 & 0 \cr
0 & 0 & 0 & 0 & I_{16} & 0 & 0 \cr
{\lambda_1\over \lambda_2} & 0 & 0 & 0 & 0 & {1\over \lambda_2} & 0
\cr
0 & 0 & -{\lambda_1\over \lambda_2}  & 0 & 0 & 0 & {
|\lambda|^2\over \lambda_2} \cr},
\eqn\aone
$$
$$
g_{\mu\nu} dx^\mu dx^\nu = -dt^2 +\lambda_2 |dz|^2,
\eqn\ekeenmetric $$
where
$$ \lambda \equiv \lambda_1 + i\lambda_2 = {i\over 2\pi} \ln {A\over z}.
\eqn\atwo
$$
$z=x^1+ix^2$ denotes
the complex coordinate labelling the two dimensional space, and
$A$ is an arbitrary complex number.
Note that this solution is ill defined for large $|z|$, since $\lambda_2$
becomes negative, forcing $\Phi$ to be complex. Thus the solution given
above should be regarded only as an approximate solution that gets
modified at large distances when $\lambda_2$ becomes of order 1 and the
theory enters the strong coupling regime. We note in passing that the
SL(2,Z) transform of these solutions with the O(8,24;Z) matrices $\Omega$
given in eq.\fourcomptwo\ will correspond to magnetically charged
strings discussed in ref.\SLTZ, winding along the $x^3$ direction.

Let us now
consider the O(8,24;Z) transformation $\omega$ that exchanges the 2nd and
31st rows and columns, and also the 9th and 32nd rows and columns:
$$\omega=\pmatrix{1 & 0 & 0 & 0 & 0 & 0 & 0 & 0\cr 0 & 0 & 0 & 0 & 0 & 0 &
1 & 0 \cr 0 & 0 & I_5 & 0 & 0 & 0 & 0 & 0 \cr 0 & 0 & 0 & 1 & 0 & 0 & 0 & 0
\cr 0 & 0 & 0 & 0 & 0 & 0 & 0 & 1 \cr 0 & 0 & 0 & 0 & 0 & I_{21} & 0 & 0 \cr
0 & 1 & 0 & 0 & 0 & 0 & 0 & 0\cr 0 & 0 & 0 & 0 & 1 & 0 & 0 & 0\cr}.
\eqn\newomega $$
This transforms the solution \aone\ to
$$\M = \pmatrix{{|\lambda|^2\over \lambda_2} & {\lambda_1\over \lambda_2}
& 0 & 0 & 0 & 0\cr
{\lambda_1\over \lambda_2} & {1\over \lambda_2} & 0 & 0 & 0 & 0 \cr
0 & 0 & I_5 & 0 & 0 & 0\cr 0 & 0 & 0 & {1\over \lambda_2} & -{\lambda_1
\over \lambda_2} & 0 \cr 0 & 0 & 0 & -{\lambda_1\over \lambda_2}
& {|\lambda|^2\over \lambda_2} & 0 \cr 0 & 0 & 0 & 0 & 0 & I_{23}},
\eqn\enewmslt $$
with $g_{\mu\nu}$ and $\lambda$ still given by eqs.\ekeenmetric\ and
\atwo\ respectively. Using the relations
\ethree, \efour, \enine\ and \eten\ between the
matrix $\M$ and the ordinary metric and antisymmetric tensor fields in
ten dimension, it can easily be checked that this corresponds to a
ten dimensional metric
of the form:
$$\pmatrix{G^\ten_{33} & G^\ten_{34}\cr G^\ten_{43} & G^\ten_{44}}
={1\over \lambda_2} \pmatrix{1 & -\lambda_1\cr -\lambda_1 & |\lambda|^2}.
\eqn\tendmetric
$$
This soluion coincides with the `stringy cosmic string solution' constructed
in ref.\SHVAFA\  near the `decompactification point' (small $|z|$).
At $z=0$ the direction 4
decompactifies, since the metric component $G^\ten_{44}$ blows up.
Note the amusing fact that in order to identify this solution to the
four dimensional solution of ref.\SHVAFA, we need to identify the coordinates
$(x^0, x^1, x^2)$, and any one of the coordinates
$(x^5, \ldots x^9)$ (let us
choose $x^5$ for definiteness) as the coordinates
of the four dimensional theory, with the string extending in the $x^5$
direction. But the coordinate $x^3$, which was the direction in which the
original fundamental string was pointing, now must be regarded as a
compact direction to start with and cannot be taken to be the direction
in which the cosmic string extends.

The solution given in eq.\enewmslt, or equivalently
eq.\tendmetric, suffers from the same kind
of problems at large distances as the solution \aone, since eigenvalues
of the metric become negative. An elegant solution to this problem was
provided in ref.\SHVAFA.
Let $j(\lambda)$ be the
SL(2,Z) invariant function of the variable $\lambda$, satisfying
$$\eqalign{
j({p\lambda + q\over r\lambda + s}) = & j(\lambda)\quad
\hbox{for} \quad  ps -rq=1, \quad p,q,r,s\in Z, \cr
\quad j(\lambda) \simeq &
e^{-2\pi i \lambda}\, \hbox{ for large }\, \lambda_2, \cr }
\eqn\eaextra $$
and that $j(\lambda)$ has no other pole anywhere else in the upper half
$\lambda$ plane.
Let us now consider a field configuration given by eq.\enewmslt,
with $\lambda(z)$ given by,
$$ j\big(\lambda(z)\big) = {A\over z} + B. \eqn\asix $$
Then, for $|z|<<|A|$, we get back eq.\atwo. This shows that for small $|z|$
the new solution agrees with the original one. However, since $\lambda(z)$
is obtained as a solution to eq.\asix, $\lambda_2$ is guaranteed to be
positive for all values of $z$. Furthermore, as $|z|\to \infty$, $\lambda$
approaches a constant value, depending on the value of $B$.
As a result, this solution is well behaved asymptotically.

This suggests that we should modify the fundamental string solution at
large distance in a similar way, so that it has good asymptotic behaviour.
In particular, we can now take the fundamental string solution to be given
by eq.\aone, with $\lambda$ given by eq.\asix\ instead of eq.\atwo.
This agrees with the original solution for small $|z|$, but differs
from it for large $|z|$, and is free from the problem of the original
solution.\foot{I wish to thank A. Shapere for a discussion on this
point.}
For $\lambda$ given as a solution of eq.\asix, $\M$ is not
a single valued function of $z$. In particular, there are closed curves
in the $z$ plane, not necessarily
enclosing the string, but the points where
$\lambda=i$ or $\exp(\pm 2\pi i/3)$, with the property
that $\lambda$ transforms by an
SL(2,Z) transformation of the form $\lambda\to (p\lambda + q)/(r\lambda+s)$
as we move around these closed curves. This induces a transformation
on $\M$ of the form:
$$ \M \to \Omega \M \Omega^T, \eqn\aeight $$
where,
$$ \Omega =
\pmatrix{ p & 0 & 0 & 0 & 0 & q
& 0 \cr
0 & I_6 & 0 & 0 & 0 & 0 & 0 \cr 0 & 0 & s & 0 & 0
& 0 & - r \cr 0 & 0 & 0 & I_6 & 0 & 0 & 0 \cr
0 & 0 & 0 & 0 & I_{16} & 0 & 0 \cr
r & 0 & 0 & 0 & 0 & s & 0
\cr
0 & 0 & - q & 0 & 0 & 0 & p \cr}. \eqn\newkk
$$
Since $\Omega$ is an O(8,24;Z) transformation, we see that the solution
obtained this way is well defined if O(8,24;Z) is a gauge symmetry of the
theory, so that we can identify field configurations related by this
transformation.

Although the solution constructed this way is well defined at this level,
it was argued in ref.\SHVAFA\ that as a result of the presence of the
orbifold points in the map from the $z$ plane to the $\lambda$ plane,
corresponding to the points $\lambda=i$ or $\lambda=\exp(\pm 2\pi i/3)$,
stringy modification of these solutions, which
preserve partial invariance of these solutions under supersymmetry
transformations, may not be possible.
\foot{The argument in ref.\SHVAFA\
was given not for the fundamental string solutions, but for the cosmic string
solutions. However, since these two solutions are related by an
O(8,24;Z) transformation,
the same argument must apply to this case as well.}
It was also shown in ref.\SHVAFA\ that there exists
12 string configurations which do not suffer from this difficulty.
This might mean that single elementary string states do not represent well
defined states, but 12 of them together do.

We end this section by mentioning that the mechanism of replacing
eq.\atwo\ by eq.\asix, that makes the solution well behaved
asymptotically, might also be responsible for the removal of
infra-red divergences alluded to in sects. 3 and 4 from other string
states as well. As stated there,
these divergences are due to the long range electric and magnetic
fields around the elementary string and soliton states.
Equivalently we can attribute these divergences to the fact that
the field $\M$ does not approach a constant value asymptotically;
this can be seen from eqs.\etwenty\ and \eextratwo. This situation
is very similar to that in the solution \atwo, where $\lambda$ does
not approach a constant value asymptotically. In particular,
$\lambda \to \lambda +1$ under $z\to e^{2\pi i} z$. Thus it is
conceivable that using the identification of field configurations
under the O(8,24;Z) symmetry, the field configurations around the
elementary string excitations and solitons can be appropriately
modified (as in eq.\asix) so that $\M$ approaches some constant value
$\M^{(0)}$ asymptotically. Such solutions will be free from the
infra-red divergence problems discussed in the previous section.

\Appendix{A}

\centerline{\bf Asymptotic Behaviour of the Periodic Array of BPS
Monopole Solutions}

In this appendix we shall discuss the asymptotic field configurations
for a periodic array of BPS monopole solutions.
We first consider a configuration of $N$ well separated monopole
solutions in four dimensions, and work in a gauge where asymptotically the
Higgs field is directed along a fixed direction (say the 3 direction) in
the isospin space everywhere except along a Dirac string singularity. The
asymptotic magnetic field is given by
$$ \vec B \simeq \sum_{I=1}^N
{\vec r - \vec r_{\zero I}\over | \vec r - \vec
r_{\zero I}|^3 } {\sigma_3\over 2}, \eqn\cone
$$
where $\vec r_{\zero I}$ is the location of the $I$th monopole. The
Bogomol'nyi equation
$$ D_i \phi = \pm B_i, \eqn\ctwo $$
for the Higgs field $\phi$ then determines the asymptotic value of the Higgs
field upto an additive constant. Choosing the $+$ sign in eq.\ctwo\ we
get,
$$ \phi \simeq \Big[ C - \sum_{I=1}^N
{ 1 \over | \vec r - \vec
r_{\zero I}| }\Big] {\sigma_3\over 2}. \eqn\cthree
$$
Here $C$ is a constant which sets the
scale of breaking of the SU(2) symmetry.

{}From this we can construct the asymptotic form for the Higgs field for a
periodic array of BPS monopoles along the three direction provided the
period is large compared to the scale of SU(2) breaking. It is given by,
$$
\phi \simeq \Big[
C - \sum_{k=-\infty}^\infty {1 \over \sqrt{ (x^1 - x^1_\zero)^2
+ (x^2 - x^2_\zero)^2 + (x^3 - x^3_\zero + 2\pi k R_{(3)})^2}}\Big]
{\sigma_3\over 2}.
\eqn\cfour
$$
The sum over $k$ in this expression has a logarithmic divergence, but
the divergene contribution is independent of the coordinates
$(x^1, x^2, x^3)$ and can
be removed by appropriately adjusting $C$. The asymptotic values of the
gauge field strengths are given by eq.\ctwo, and are finite. This way we
can construct a soliton solution in the three dimensional string theory
from periodic array of BPS monopoles in four dimensions. The solution,
however, suffers from the usual problem of logarithmic growth of the
scalar field $\phi$ at large distance, and the associated problem of
infinite energy. In particular, if we define,
$$ \rho = \sqrt {(x^1 - x^1_\zero)^2 + (x^2 - x^2_\zero)^2} \, ,\eqn\ceight $$
then for large $\rho$,
$$ \phi\simeq {1\over 2 \pi R_{(3)}} (\ln\rho) \sigma_3.
\eqn\easympone $$

The $N$-soliton solution in three dimensions can be constructed in an
identical manner from $N$ different periodic arrays of BPS monopoles, and
has  the following asymptotic form of the Higgs field when the arrays are
well separated from each other:
$$
\phi \simeq \Big[ C - \sum_{I=1}^N
\sum_{k=-\infty}^\infty {1 \over \sqrt{ (x^1 - x^1_{\zero I})^2
+ (x^2 - x^2_{\zero I})^2 + (x^3 - x^3_{\zero I} + 2\pi k R_{(3)})^2}}
\Big] {\sigma_3\over 2}.
\eqn\cfive
$$

Finally, we would like to point out that although these solutions are not
independent of the coordinate $x^3$, and hence are not vortex
solutions in the conventional sense, the dependence of these solutions on
$x^3$ decreases exponentially as we move away from the core. To
see this, let us consider a single array of monopoles. In this case, the
asymptotic Higgs field is given by eq.\cfour, and hence,
$$\eqalign{
{\p\phi\over\p x^3} \simeq & \sum_{k=-\infty}^\infty
{(x^3-x^3_\zero+2\pi k R_{(3)}) \over \big( \rho^2
+ (x^3 - x^3_\zero + 2\pi k R_{(3)})^2\big)^{3/2}}
\cdot {\sigma_3\over 2} \cr
=& \ointop {dw\over 2i} \cot (\pi w)
{(x^3-x^3_\zero+2\pi w R_{(3)}) \over \big( \rho^2
+ (x^3 - x^3_\zero + 2\pi w R_{(3)})^2\big)^{3/2}}
\cdot {\sigma_3\over 2}\, , \cr}
\eqn\csix
$$
where the contour of integration over $w$ encloses the real axis.
We can now express this as the sum of two contours, $C_1$ and
$C_2$, given by,
$$\eqalign{
w = & y - {1 \over 2\pi R_{(3)}}\{ (x^3 - x^3_\zero) -
i (\rho -\epsilon) \}
\quad \quad \hbox{on } C_1, \cr
w = & - y - {1 \over 2\pi R_{(3)}}\{ (x^3 - x^3_\zero)
+ i (\rho -\epsilon) \}
\quad \quad \hbox{on } C_2, \cr}
\quad \quad -\infty < y < \infty, \quad y \, \, {\rm real},
\eqn\cseven
$$
where $\epsilon$ is a small but fixed number.
Eq.\csix\ may then be rewritten as
$$\eqalign{
{\p\phi\over \p x^3} \simeq & \int_{-\infty}^\infty {dy\over 2i}
{ 2\pi R_{(3)} y + i (\rho-\epsilon) \over
\big(\{2\pi R_{(3)} y + i (2\rho -
\epsilon)\} \{ 2\pi R_{(3)} y -i\epsilon\}\big)^{3/2} } \cr
& \times \Big[
\cot\{ { i(\rho -\epsilon)
- (x^3 - x^3_\zero) \over 2 R_{(3)}} + \pi y \}
- \cot\{ { i(\rho -\epsilon)
+ (x^3 - x^3_\zero) \over 2 R_{(3)}}+\pi y \} \Big]
{\sigma_3\over 2},\cr
}
\eqn\cnine
$$
where the factor $\big(\{ 2\pi R_{(3)} y + i (2\rho - \epsilon)\}
\{ 2\pi R_{(3)} y -i\epsilon\}\big)^{3/2}$ in the denominator
is to be taken to be positive for $y=0$. This defines this term on the
real $y$ axis completely since there is no branch point on this axis.
The term inside the square bracket is of order
$e^{-(\rho -\epsilon)/R_{(3)}}$
for large $\rho$. The rest of the integrand is finite and falls off as
$y^{-2}$ for large $|y|$. This shows that
$(\p\phi/\p x^3)$ falls off exponentially as we move away from the soliton
core.

We end with the note that our analysis so far does not show that
the solutions representing periodic arrays of BPS monopoles, constructed
this way, are non-singular everywhere in space. As we shall see in
appendix B, these solutions give rise to well defined scattering
functions, which might be an indication that the solutions are indeed
finite everywhere in space. In particular, if the solution is singular
at some point, then we might expect the scattering function
associated with a line passing through that point to be ill defined.
Even if it turns out that the solutions
have singularity at some point(s) in space,
we might hope that stringy corrections will remove the singularity,
or that string propagation in the background provided by these
solutions is well behaved, even if the solutions themselves are
singular.

\Appendix{B}

\centerline{\bf Moduli Space of
Periodic Arrays of BPS Multi-monopole Solutions}

In this appendix we study the moduli space of periodic array of BPS
multi-monopole solutions in (3+1) dimensions. We start with the
observation\ATHIT\ that moduli space of $N$ monopole solutions is identical
to the space of rational functions $S(v)=P(v)/Q(v)$, where $P(v)$ is a
polynomial in $v$ of degree $\le N-1$,
$Q(v)$ is a polynomial in $v$ of degree
$N$, and the zeroes of $Q(v)$ are distinct from the zeroes of $P(v)$.
For $N$ well separated monopoles the zeroes of $Q(v)$ are all
distinct, and $S(v)$ has the form:
$$ S(v) = \sum_{I=1}^N {\lambda_I \over v - v_I}. \eqn \dtwo $$
$S(v)$ has the interpretation of being the scattering function associated
with the the differential operator $\p_t + i A_t -\phi$ on a line parallel
to the $x^2$ axis, where $t$ denotes length measured along the line,
and $v=x^3+i x^1$ labels the point of intersection of the line with the
$1-3$ plane.\foot{Instead of choosing the lines to be parallel to the
$x^2$ axis, we could have chosen any other set of parallel lines. For a
detailed discussion of this see Ref.\ATHIT.} Thus, a periodic array of
monopoles along the 3 axis with period $2\pi R_{(3)}$  will be
described by a scattering function of the form:
$$  S(u) = \sum_{n=-\infty}^\infty {\lambda\over u - u_0 - 2\pi n}
={\lambda\over 2} \cot[{1\over 2}(u-u_0)]. \eqn\edthree
$$
where $u\equiv (x^3 + i x^1)/R_{(3)}$.
It is also straightforward to write down the scattering function
$S(u)$ for
$N$ well separated arrays of BPS monopoles. This is given by,
$$  S(u)
={1\over 2} \sum_{I=1}^N \lambda_I \cot[{1\over 2}(u-u_{I})],
\quad u_{I}\ne u_{J} \hbox{ for } I\ne J. \eqn\dfour
$$
This function is parametrized by $2N$ complex variables $\lambda_I$ and
$u_{I}$.

The function $S(u)$ given in eq.\dfour\ is characterized by the fact that
it is periodic in $Re (u)$ with period $2\pi$,
it has $N$ distinct poles in the
interval $0\le Re (u) < 2\pi $, and that $\lim_{u\to i\infty} S(u)
= - \lim_{u\to -i\infty} S(u)$, but neither of these limits vanish.
A more well behaved scattering function, one that vanishes for $Im (u)
\to \pm \infty$, may be obtained if we consider quasi-periodic arrays of
monopoles which pick up a U(1) phase $e^{2\pi i\alpha}$ under translation
by one period.\foot{This corresponds to introducing a Wilson line along
the 3 direction\GAUNTHAR.}
A single array of this kind is described by a scattering
function
$$ S(u) = \sum_{n=-\infty}^\infty {\lambda e^{2\pi i\alpha n}\over
u-u_0- 2\pi n}
= {\lambda\over 2} {e^{i(\alpha-{1\over 2})(u-u_0)}\over \sin
[{1\over 2} (u-u_0)]},
\, \, \quad \hbox{ for } \, 0<\alpha<1.
\eqn\quasione $$
(Up to an overall normalization factor, this result may also be derived
by starting from the configuration of cyclically symmetric monopoles
given in ref.\MURMAN\ and then taking the limit where the radius of
the circle becomes large.)
A configuration of multiple well separated  arrays
is described by the scattering
function:
$$ S(u) = {1\over 2} \sum_{I=1}^N \lambda_I
{e^{i(\alpha-{1\over 2})(u-u_{I})}\over \sin[{1\over 2} (u-u_{I})]},
\, \, \quad \hbox{ for } \, 0<\alpha<1.
\eqn\quasitwo $$
Note that $S(u)$ vanishes for $Im(u)\to \pm\infty$.
This fall-off may be attributed to the cancellation between the contributions
from different monopoles in the array due to their phase difference.
In analogy with the BPS monopole solutions in four dimensions, we now
speculate that
for the most general configuration of $N$ quasi-periodic arrays
(not necessarily well-separated) of this kind,
the scattering function $S(u)$ satisfies the following criteria:
\item{1.} $S(u+2\pi) = e^{2\pi i\alpha} S(u).$

\item{2.} $\lim_{u\to \pm i\infty} S(u) = 0.$

\item{3.} $S(u)$ has $N$ (not necessarily distinct) poles in the interval
$0\le Re (u) < 2\pi$.

A general expression for such a function is given by
$$
S(u) = {e^{i (\alpha+N/2)u} \sum_{J=1}^N a_J e^{-iJu} \over \prod_{I=1}^N
\sin[{1\over 2} (u - u_I)]}\, ,
\eqn\gensu $$
with the constraint
$$ \prod_{I=1}^N \sum_{J=1}^N a_J e^{-iJ u_I} \ne 0.
\eqn\suconstraint
$$
Eq.\suconstraint\ guarantees that the poles in $S(u)$ coming from the
zeroes in the denominator are not cancelled by the zeroes of the
numerator.
We shall denote by $\M_N$ the $4N$ real dimensional space parametrized
by $a_I\in C$ and $u_I\in S^1\times R$, satisfying \suconstraint.

There is a natural action of the group $R^2\times S^1\times S^1$ on the
function $S(u)$ and hence on the space $\M_N$.
This group action is identical to that for $N$ monopole
solutions in four dimensions, with the only difference that the
translation in the third direction is now represented by a compact
coordinate. If $(X^1, X^2, X^3, \Theta)$ denote
the parameters of the group, with $X^3$ and $\Theta$ periodic
parameters with period $2\pi$, then under the action of this group,
$$ S(u) \to e^{X^2 +  i\Theta} S(u- X^3 - iX^1).
\eqn\dten $$
Physically, $X^1$, $X^2$ and $X^3$ denote translations in the
1, 2 and 3 directions respectively, while $\Theta$
represents a change of the U(1) phase.
Using these symmetry transformations, we can locally `gauge fix' the
coordinates $a_I$, $u_I$ of the full moduli space $\M_N$ to
$\wt a_I$, $\wt u_I$, satisfying,
$$
\sum_{I=1}^N \wt u_I =  0 \eqn\gfixone $$
and,
$$
\prod_{I=1}^N \sum_{J=1}^N \wt a_J e^{-iJ\wt u_I} =  1\, .
\eqn\gfixtwo $$
We shall denote by $\M_N^0$ the $4(N-1)$ dimensional space spanned
by the coordinates $\wt a_J\in C$ and $\wt u_I \in S^1\times R$,
satisfying eqs.\gfixone, \gfixtwo. Then locally the full moduli
space $\M_N$ parametrized by the coordinates $a_I$ and $u_I$ subject
to the constraint \suconstraint\ can be described the coordinates
$(X^1, X^2, X^3, \Theta)$ of $R^2\times S^1 \times S^1$ and those of
$\M_N^0$ through the relations:
$$
a_J = \wt a_J e^{X^2 + i\Theta} e^{i J (X^3 + i X^1)} \eqn\gunfixone
$$
and,
$$
u_I = \wt u_I + X^3 + i X^1\, . \eqn\gunfixtwo
$$
Thus locally the full moduli space $\M_N$ has the structure of
$R^2\times S^1 \times S^1 \times \M_N^0$. Globally, however,
we need a $Z_N\times Z_N$ identification of the points in this
space to recover the correct moduli space.
This is seen by noting that the following $Z_N\times Z_N$
transformations in $R^2\times S^1\times S^1\times \M_N^0$ leave
the original parameters $a_I$ and $u_I$ unchanged (up to shift
of $u_I$ by multiples of $2\pi$):
$$\eqalign{
\wt u_I\to & \wt u_I +{2\pi\over N} \quad \hbox{for} \quad 1\le I
\le N-1, \quad \wt u_N \to \wt u_N +{2\pi \over N} - 2\pi, \cr
\quad \wt a_J \to & \wt a_J e^{2\pi i J\over N}\, ,
\quad X^3\to  X^3 -{2\pi\over N},
\cr}
\eqn\identone $$
and,
$$ \wt a_J \to \wt a_J e^{2\pi i\over N}, \quad \Theta\to \Theta -
{2\pi \over N}\, . \eqn\identtwo $$
This establishes the structure \structure\
of the moduli space of multiple array of BPS monopole solutions.
For well separated arrays, the transformation \identone\ corresponds
to translating the $N$th array by a distance $2\pi$ along the negative
3-axis relative to the other arrays, accompanied by an overall shift
of all the arrays by a distance $2\pi/N$ along the positive 3-axis.

\Appendix{C}

\centerline{\bf Metric on $\M_N$}

We shall begin by discussing a subtlety involving the overall U(1)
phase $\Theta$. A change of $\Theta$ by $\delta\Theta$ corresponds to
an infinitesimal gauge transformation of the solution with matrix
valued parameter $\Lambda(x^1, x^2, x^3)\equiv \sum_{a=1}^3 \Lambda^a
\sigma^a/2$ which approaches the value
$\delta\Theta\, \sigma_3/2$ far away from the soliton core.
In order to see what kind of deformation of the solution it
corresponds to, we run into a problem. In we demand as in ref.\ATHIT\
that the
deformation generated by the gauge transformation parameter $\Lambda$
is orthogonal to all the pure gauge deformations\foot{In this case
orthogonality is only in a formal sense, since the proof of orthogonality
given in ref.\ATHIT\ involves an integration by parts, and the
relevant boundary terms do not vanish in this case.} then $\Lambda$
must be proportional to the background Higgs field $\phi$\ATHIT:
$$ \Lambda(x^1, x^2, x^3) =
\delta\wh\Theta \, \phi(x^1, x^2, x^3)\, , \eqn\koon $$
where $\delta \wh\Theta$ is a free parameter.
Eqs.\cfive\ and \koon\ shows that asymptotically $\Lambda\simeq
(N/2\pi R_{(3)})\delta\wh\Theta (\ln \rho) \sigma_3$.
Comparing this with the asymptotic
form of $\Lambda$ generated by the deformation of the parameter $\Theta$,
we see that the deformation induced by $\delta\wh\Theta$ is not
compatible with the deformation induced by $\delta\Theta$. Furthermore,
as we shall see later, a deformation induced by $\Lambda$ given
in eq.\koon (as well as the deformations induced by the parameters
$X^1$ and $X^2$) are not normalizable. All these difficulties may be
avoided by introducing an explicit infra-red cut-off at some length
scale $L$.
In this case all the deformations will be normalizable, and
the parameter $\Lambda$ given by eq.\koon\
approaches the value $(N/2\pi R_{(3)})\delta\wh\Theta(\ln L) \sigma_3$
asymptotically.
Thus now $\wh\Theta$ and $\Theta$ can be related as $\Theta
=N\wh\Theta \ln L/\pi R_{(3)}$.

Such a regularization
may be provided, for example, by keeping the extent of
the array in the third direction to be large but finite, instead of being
strictly periodic and infinite.
If the array stops after $L$ repetitions, then locally $\M_N$ can
be approximately identified to a $4N$ dimensional subspace of the
$4NL$ dimensional moduli space $M_{NL}$ of $NL$ monopoles in
$3+1$ dimensions. Equivalently, one could consider a $Z_L$ symmetric
cyclic arrangement\MURMAN\
of $NL$ monopoles around a circle of radius $LR_{(3)}$
in the $x^1-x^3$ plane,
each monopole being a finite distance away from the circle.
In the $L\to \infty$ limit this represents $N$ periodic arrays of
monopoles with periodicity $2\pi R_{(3)}$.

Using arguments similar to those in ref.\ATHIT\ one can show
that the vector fields $\p/\p X^1$, $\p/\p X^2$, $\p/\p X^3$ and
$\p/\p \wh\Theta$ are covariantly constant in
the full moduli space.
As a result, the metric on the full moduli space $\M_N$ factorizes into a
direct sum of a constant metric on $R^2\times S^1\times S^1$ and a
metric on $\M_N^0$. We shall now show that although the constant metric
on $R^2\times S^1\times S^1$ depends on the infra-red regulator $L$,
the metric on $\M_N^0$ is in fact independent of $L$ and hence the
notion of the existence of harmonic forms on $\M_N^0$ is well defined.
We first use eq.\cfive\ to express the
asymptotic value of the Higgs field as
$$ \phi=\{\wt C + {N\over \pi R_{(3)}} \ln |\vec \rho -\vec R|
+ O({1\over \rho^2})\} {\sigma_3 \over 2}
+ \sum_{i=1}^2 \sigma_i f_i(\vec \rho),
\eqn\appnone
$$
where,\foot{Note that the center of mass coordinates $X^1$, $X^2$, $X^3$
have been scaled by $R_{(3)}$ so that $X^3$ has periodicity $2\pi$. This
notation is consistent with the one used in appendix B.}
$$ \vec R \equiv (R_{(3)} X^1, R_{(3)} X^2), \quad
\quad X^i\equiv {1\over N R_{(3)}} \sum_{I=1}^N
x^i_{\zero I}, \quad \quad i=1,2,
\eqn\appntwo $$
$$ \vec \rho = (x^1, x^2), \quad \quad
\rho^2 =\sum_{i=1}^2 (x^i)^2. \eqn\appnthree $$
Here $\wt C$ is a finite constant, $\sigma_i$ are Pauli matrices,
and $f_i(\vec \rho)$ are some functions of $\vec \rho$.
Although we do not know
the explicit form of $f_i(\vec \rho)$, we know that they fall off
exponentially for $\rho>> \wt C^{-1}$, {\it i.e.}
outside the core of the
soliton.
The moduli space of soliton solutions corresponds to the space of solutions
for fixed  $N$ and $\wt C$, i.e. fixed asymptotic value of the Higgs (and
the other) field(s). It is easily seen from eq.\appnone\ that the overall
translation in the $x^1-x^2$ plane corresponds to a shift of $\vec R$.
Since a change in $\vec R$ gives $\delta\phi\propto \delta\vec R
\cdot \vec \rho/ \rho^2$ asymptotically, the zero modes associated
with these deformations are clearly non-normalizable.
This shows that the zero modes associated with the
$R^2$ factor in eq.\structure\ are not normalizable. But this analysis
also shows that all other
deformations of the solution which keep $\vec R$ (and also $\wt C$)
fixed give rise to square
integrable $\delta\phi$.

Thus in order to test the normalizability of the zero modes corresponding to
these other deformations we need to study whether the
deformations of the gauge fields induced by these deformations are also
square integrable. In this case the gauge fields in fact are not independent,
but are related to $\phi$ (up to a gauge transformation) by the Bogomol'nyi
equation
\ctwo. Using this equation,
one can verify that the deformations that keep
$\wt C$, $N$ and $\vec R$ in eq.\appnone\ fixed induces a
gauge field deformations for which $\delta F_{ij}\sim 1/\rho^3$, and hence
$\delta A_i\sim 1/\rho^2$ asymptotically. Such deformations are clearly
square integrable. However we also have deformations of the
solution induced by the gauge transformations with
parameter $\Lambda$ given in eq.\koon.
This induces an asymptotic gauge field deformation of the
form:
$$ \delta A_i \propto {N\over 2\pi R_{(3)}} \delta\wh\Theta
\sigma_3 {x^i\over \rho^2}. \eqn\appnfour $$
This deformation is clearly not square integrable.

Thus we see that the deformations associated with the parameters
$X^1$, $X^2$ and $\wh\Theta$ are not square integrable, but
all other deformations are square integrable.
In particular, all deformations
associated with the coordinates labelling $\M_N^0$ are square
integrable. Thus the metric on $\M_N^0$ is finite and hence is
independent of the infrared cut-off $L$. On the other hand the metric
components $g_{X^1 X^1}$, $g_{X^2 X^2}$ and $g_{\wh\Theta \wh\Theta}$
are all of order $\ln L$. Using the relation $\Theta=N\wh\Theta\ln L
/\pi R_{(3)}$,
we see that $g_{\Theta\Theta}$ is of order $(\ln L)^{-1}$.

Even though the metric on $\M_N^0$ is finite,
unlike in the case of four dimensional theory, $\M_N^0$
does not become flat in the limit
when the solitons are far from each other. This is related to the
non-normalizability of the zero modes describing the dynamics of single
isolated solitons.
We shall now derive the form of the metric on the
two soliton moduli space in the limit where the two solitons are widely
separated. Let $X^i$ ($1\le i\le 3$) denote the center of
mass coordinates labelling the space $R^2\times S^1$, $\Theta$
be the overall U(1) phase labelling the second $S^1$, and $\wt x^i$,
$\wt\theta$ be the relative coordinates labelling the space $\M_2^0$. We
shall normalize the coordinate
$\wt \theta$ such that it has periodicity $2\pi$.
Following the analysis given at the end of appendix A, we can show
that a change in the solution due to a change in
$\wt x^3$ falls off sufficiently fast asymptotically, hence $\wt g_{33}$ is
expected to be finite asymptotically. On the other hand, the asymptotic value
of the Higgs field as a function of $\vec R\equiv (R_{(3)} X^1,
R_{(3)} X^2)$
and $\vec{\wt \rho} \equiv (\wt x^1, \wt x^2)$ is given by,
$$ \phi \simeq \Big \{ \wt C + {1\over \pi R_{(3)}} \ln |\vec \rho -
(\vec R + \vec{\wt \rho})|
+ {1\over \pi R_{(3)}}
\ln |(\vec \rho - (\vec R - \vec{\wt \rho})| \Big\}\,{\sigma_3\over 2}
\, . \eqn\metricone
$$
If $\delta\phi$ denotes
the change of $\phi$ under a variation $\delta\vec{\wt \rho}$ of
$\vec{\wt \rho}$
for fixed $\vec R$, then it can easily be verified
that,
$$ \int d^2 \rho |\delta\phi|^2 \propto \ln{\wt \rho} |
\delta\vec{\wt \rho}|^2,
\eqn\metrictwo $$
with the major contribution to the integral coming from two regions:
$|\vec \rho -\vec R -\vec{\wt \rho}|<< \wt \rho$, and
$|\vec \rho -\vec R +\vec{\wt \rho}|
<< \wt \rho$.  A similar contribution is obtained from the computation of
$\int d^2 \rho |\delta A_i|^2$.
This gives $\wt g_{ij} \propto \ln \wt \rho \,
\delta_{ij}$ for $1\le i,j\le 2$.

Finally we need to analyze $\wt g_{\theta\theta}$. In this case we proceed
in a somewhat indirect manner. Let $\wt p_\theta$ be the
momentum conjugate to $\wt\theta$.
$\wt p_\theta$ is quantized in integer units since $\wt\theta$ has been
normalized to have periodicity $2\pi$.
For large separation between the two solitons, $\wt p_\theta$
has the interpretation as the difference in the electric charges carried
by the two solitons.
Now, as has been argued at the end of Sec.4, if $\wt p_\theta\ne 0$,
i.e. if the two
solitons carry different amount of electric charges, then
the attractive force between them proportional
to $1/\wt \rho$ due to the
higgs field wins over the repulsive force due to the electromagnetic fields.
In the weak coupling limit, this net
force is proportional to $(\wt p_\theta)^2
/\wt \rho$.
The net result is a contribution to the potential energy (and hence to
the Hamiltonian) proportional to
$(\wt p_\theta)^2 \ln\wt \rho$.
This, in turn, means that the Lagrangian
of the system must contain a term proportional to $(\ln \wt \rho)^{-1}
(d \wt\theta/dt)^2$. Thus $\wt g_{\theta\theta}
\propto (\ln \wt \rho)^{-1}$.

Combining the above results, we get the following asymptotic form of
different components of the metric
$$\wt g_{ij}\simeq C_1\ln \wt \rho \, \delta_{ij},
\quad \wt g_{33}\simeq C_2, \quad
\wt g_{\theta\theta}\simeq C_3 (\ln \wt \rho)^{-1},
\eqn\metricfin
$$
where $C_i$ are constants.

\ack I wish to thank J. Cohn, A. Dabholkar,
J. Harvey, J. Schwarz, A. Shapere, S. Shenker, C. Vafa and E. Witten
for useful discussions. I would also like to acknowledge the hospitality
of the Department of Physics at Rutgers university, where
part of this work was done.

\refout

\end